\documentclass[11pt,a4paper]{amsart}
\usepackage[margin=1.1in]{geometry}
\usepackage{amsmath,amssymb,amsthm,mathtools}
\usepackage{xcolor}
\usepackage[colorlinks=true,linkcolor=blue!60!black,citecolor=green!40!black,urlcolor=blue!60!black]{hyperref}
\usepackage{enumitem}
\usepackage{booktabs}
\usepackage[expansion=false]{microtype}
\usepackage{epigraph}
\theoremstyle{plain}
\newtheorem{theorem}{Theorem}[section]
\newtheorem{proposition}[theorem]{Proposition}
\newtheorem{lemma}[theorem]{Lemma}
\newtheorem{corollary}[theorem]{Corollary}
\newtheorem{conjecture}[theorem]{Conjecture}
\newtheorem*{thmA}{Theorem A}

\theoremstyle{definition}
\newtheorem{definition}[theorem]{Definition}

\newtheorem{construction}[theorem]{Construction}
\theoremstyle{remark}
\newtheorem{remark}[theorem]{Remark}

\newcommand{\Z}{\mathbb{Z}}
\newcommand{\Q}{\mathbb{Q}}
\newcommand{\C}{\mathbb{C}}
\newcommand{\F}{\mathbb{F}}
\newcommand{\cA}{\mathcal{A}}
\newcommand{\cB}{\mathcal{B}}
\newcommand{\cC}{\mathcal{C}}
\newcommand{\cD}{\mathcal{D}}
\newcommand{\cE}{\mathcal{E}}
\newcommand{\cS}{\mathcal{S}}
\newcommand{\cM}{\mathcal{M}}

\newcommand{\Mat}{\operatorname{Mat}}
\newcommand{\Br}{\operatorname{Br}}
\newcommand{\supp}{\operatorname{supp}}
\newcommand{\ind}{\operatorname{ind}}
\newcommand{\Ad}{\operatorname{Ad}}
\newcommand{\id}{\mathrm{id}}
\newcommand{\mul}{\operatorname{mul}}
\newcommand{\Qpos}{\Q_{>0}}
\newcommand{\QCA}{\mathcal{Q}}
\newcommand{\Circ}{\mathcal{C}}

\newcommand{\Pau}{\Pi}

\DeclareMathOperator{\coker}{coker}
\DeclareMathOperator{\width}{width}

\title[Three-dimensional QCA that are not stably circuits]{Three-dimensional quantum cellular automata\\ that are not stably circuits}
\author{Bowen Yang}
\address{Harvard University, Cambridge, MA 02138, USA}
\email{bowen\_yang@g.harvard.edu}

\begin{document}
\begin{abstract}
We prove that the three-dimensional chiral $\Z_p$ Clifford quantum cellular automata of Haah and of Sun--Yang--Wang--Tantivasadakarn--Chen are, for every odd prime $p$, not finite-depth quantum circuits, even after tensoring with arbitrary ancillary degrees of freedom and composing with translations. Previously such statements were known only inside the Clifford category. As a consequence, the group of three-dimensional QCA modulo stabilized circuits and translations is not finitely generated. The key ingredient is an invariant of invertible subalgebras, which we conjecture to be the square of the total quantum dimension of the hosted anyon theory, modulo rational squares.
\end{abstract}
\maketitle
\epigraph{Two is celebrated as the only even prime, which in some sense makes it the oddest prime of all.}{--- John H. Conway and Richard K. Guy, \textit{The Book of Numbers}}
\setcounter{tocdepth}{1}
\tableofcontents
\section{Introduction}\label{sec:intro}

A quantum cellular automaton (QCA) on a lattice of finite-dimensional quantum systems is a $*$-automorphism of the algebra of local operators which moves every operator by a bounded distance. The simplest QCA are finite-depth quantum circuits and translations, and the central question of the subject is which QCA are, up to these trivial ones, genuinely new. In one dimension the answer is given by the GNVW index \cite{GNVW}. In two dimensions every QCA is a circuit composed with a translation \cite{FH}. In three dimensions Haah, Fidkowski and Hastings \cite{HFH} constructed a QCA that disentangles the three-fermion Walker--Wang model and showed that it cannot be a circuit \emph{if} the three-fermion anyon theory admits no commuting-projector realization in two dimensions; this is widely believed but not known. Haah \cite{Haah-Clifford} then showed that the group of three-dimensional Clifford QCA modulo Clifford circuits and translations contains a copy of the Witt group of quadratic forms over $\F_p$, and he gave explicit nontrivial examples; Sun, Yang, Wang, Tantivasadakarn and Chen \cite{SYWTC} constructed Clifford QCA realizing all these Witt classes. In all of these results, nontriviality is established only against \emph{Clifford} circuits. Whether any of these three-dimensional QCA is nontrivial as a QCA has remained open.

In this paper we settle this question for the chiral $\Z_p$ QCA, for every odd prime $p$. To state the result, write $\cA(X,p)$ for the local operator algebra of a spin system on a countable set $X$ with  local dimensions $p\colon X\to\Z_{\ge1}$ (Section~\ref{sec:prelim}). For an odd prime $p$ and $k\in\F_p^\times$, let $\cB_{p,k}\subseteq\cA(\Z^2,p^2)$ be the translation-invariant Pauli invertible subalgebra of \cite[\S V.B]{SYWTC}. For $p=3$ it hosts a chiral $\Z_3$ anyon theory, like the subalgebra of \cite[\S5]{Haah-ISA}; the two subalgebras are different, but they are related by explicit locality-preserving isomorphisms and a reflection of the plane, and Theorem~A applies to both (Remark~\ref{rem:haah-z3}). Let $\alpha_{p,k}$ be the corresponding QCA of $\cB_{p,k}$ on $\cA(\Z^3,p^2)$ (Section~\ref{sec:isa}). 

\begin{thmA}[Theorem~\ref{thm:main}]
Let $p$ be an odd prime and $k\in\F_p^\times$. The Brauer class $[\cB_{p,k}]\in\Br_2$ is nonzero, and the QCA $\alpha_{p,k}$ is not stably a circuit.
\end{thmA}

Since $I(\cB_{p,k})=[p]$ for every odd prime $p$, the group of three-dimensional QCA modulo stabilized circuits and translations is not finitely generated. In fact $\Br_2$ contains a subgroup isomorphic to $\bigoplus_{p\equiv1\,(4)}\Z/2$ (Corollary~\ref{cor:inf-gen}).

The paper is written so as to be self-contained given the results of \cite{Haah-ISA} and \cite{GNVW}. The proof is a combination of three ingredients.

\subsubsection*{Invertible subalgebras and the reduction to dimension two}
Haah \cite{Haah-ISA} introduced \emph{invertible subalgebras}: subalgebras $\cB$ of a local operator algebra $\cA$ such that $\cA$ locally factors as $\cB\otimes\cC$ with $\cC$ the commutant of $\cB$. Their stable equivalence classes form an abelian group $\Br_d$. Haah proved that the \emph{boundary algebra} of a $(d+1)$-dimensional QCA $\alpha$ along a hyperplane is a $d$-dimensional invertible subalgebra whose class $\partial\alpha\in\Br_d$ is invariant under blending. Moreover, every class is the boundary class of a \emph{pumping QCA} $\alpha_\cB$, and $\partial$ identifies $\Br_d$ with the group of $(d+1)$-dimensional QCA modulo blending equivalence and shifts. Ji and the author showed that, for $d\ge1$, this coincides with the group of $(d+1)$-dimensional QCA modulo stabilized circuits \cite{JY-spectrum}. Then, \cite{Ludewig} introduced \emph{Azumaya nets}, an intrinsic definition of invertible subalgebras, and showed that QCA form the degree-zero part of a coarse homology theory built from the $K$-theory of Azumaya nets, recovering the $\Omega$-spectrum of \cite{JY-spectrum}. Theorem A is therefore reduced to showing that $[\cB_{p,k}]\ne0$ in $\Br_2$.

\subsubsection*{The half-plane invariant}
Our tool is an invariant of two-dimensional invertible subalgebras that lives in $\Qpos/\Qpos^2$. Let $\cB\subseteq\cA=\cA(\Z^2,p)$ be invertible with commutant $\cC$. Two copies of the system decompose as $\cA^{\otimes2}=\cB^{(1)}\cB^{(2)}\cC^{(1)}\cC^{(2)}$ and the automorphism $\tau_\cB$ that exchanges the two copies of $\cB$ while fixing the two copies of $\cC$ is an involutive QCA, called the \emph{Brauer swap}. A \emph{truncation} of $\tau_\cB$ to the upper half-plane is a QCA $t$ which agrees with $\tau_\cB$ far above the line $L=\{x_2=0\}$ and with the identity far below it. Since $\tau_\cB^2=1$, the QCA $t^2$ is supported near $L$, so it is a one-dimensional QCA and has a GNVW index $\ind(t^2)\in\Qpos$. We prove (Proposition~\ref{prop:well-defined}) that, when such truncations exist with uniform bounds, the class
\[
I(\cB):=\big[\ind(t^2)\big]\in\Qpos/\Qpos^2
\]
is well-defined and vanishes when $\cB$ is stably trivial (Proposition~\ref{prop:Brauer-invariance}). 

\subsubsection*{The computation}
For a Pauli invertible subalgebra the Brauer swap is a Clifford QCA, given by an explicit symplectic matrix over Laurent polynomials (for $\cB_{p,k}$ it is displayed in \eqref{eq:tau-closed}). We construct an explicit truncation of the Brauer swap of $\cB_{p,k}$ with $\ind(t^2)=p^{-1}$, so that $I(\cB_{p,k})=[p]\ne0$. The same method computes $I$ for a general class of translation-invariant Pauli invertible subalgebras (Corollary~\ref{cor:parity}).

\subsubsection*{Comments}
(1) After the first version of this paper was posted, the author became aware that Jones, Sopenko, and Thorngren~\cite{jones2026quantum} use the action of the symmetric group \(S_N\) on \(N\) copies to establish the nontriviality of certain one-dimensional invertible quasi-local algebras. This yields nontrivial two-dimensional QCAs on systems with infinite-dimensional on-site Hilbert spaces. These are genuinely new examples, going beyond what is possible with finite-dimensional on-site Hilbert spaces. The \(\mathbb{Z}/2\) Brauer swap considered here can be viewed as a related construction in two dimensions. More broadly, permutation symmetry has also been used to construct invariants of invertible phases in~\cite{sopenko2024index}.

(2) The invariant $I(\cB)$ is a close relative of invariants that have appeared in the physics literature. In the language of \cite{SZJL} it is the obstruction to on-siteability of the $\Z/2$-symmetry $\tau_\cB$. In the language of \cite{JY-obstruction} it is the degree-two universal obstruction class of the QCA representation $\Z/2\to\QCA(\Z^2)$, $\sigma\mapsto\tau_\cB$, computed through the truncation cocycle of \cite[App.~A]{JY-obstruction}. 

(3) Theorem~\ref{thm:main} is, to our knowledge, the first unconditional proof that a three-dimensional QCA is not stably a circuit. The nontriviality of the three-fermion QCA of~\cite{HFH} was previously announced in~\cite{Haah2026FermionicQCATalk}. The triviality of two-dimensional fermionic QCAs, was subsequently established in~\cite{kwan2026fermionic}.

(4) We do not expect our method to say anything about the three-fermion QCA (Remark~\ref{rem:3F}). We explain in Remarks~\ref{rem:why} and~\ref{rem:anyons} why the invariant appears to count the anyons of the theory hosted by $\cB$ modulo squares, which would account for the dichotomy between $\Z_p$ ($p$ anyons) and the three-fermion theory ($4$ anyons). For translation-invariant Clifford QCA on $\Z^3$ with qudits of odd prime dimension, this count follows from Theorem~\ref{thm:computation} and the classification of \cite{Haah-Clifford-cat}.

(5) The invariant $I$ is defined for every invertible subalgebra, Clifford or not, and it depends only on the Brauer class (Theorem~\ref{thm:Brauer-invariance}). The Clifford structure is used only to compute $I$ explicitly, for $\cB_{p,k}$ and in Corollary~\ref{cor:parity}. We leave the computation of $I$ for many other three-dimensional QCA to future work. See, for example, \cite{sun2026non,shirley2022three,chen2026invertible, fidkowski2024pumping, fidkowski2025quantum} and the references therein.

\subsubsection*{Organization}
Section~\ref{sec:prelim} fixes conventions on spin systems, QCA and circuits and recalls the GNVW index. Section~\ref{sec:isa} recalls Haah's invertible subalgebras, boundary algebras and pumping QCA, and proves the reduction to $\Br_2$. Section~\ref{sec:invariant} constructs the half-plane invariant and proves its properties. Section~\ref{sec:pauli} collects the facts about Clifford QCA that we need, including the index of separated Clifford QCA of~\cite{yang2026categorifying}. Section~\ref{sec:examples} defines the chiral $\Z_p$ subalgebras, computes the invariant, records a general cokernel-parity formula and proves Theorem A. Section~\ref{sec:discussion} discusses the relation with anyons and formulates conjectures.

\subsubsection*{Notation} We use additive notation for the quotient abelian group $\Qpos/\Qpos^2$. Tuples of vectors such as $(v_X,v_Z)$, $(a,b)$ or $(v_1,v_2)$ are often written as rows for typographical convenience, but whenever a matrix acts on them they are to be read as column vectors, $(v_X,v_Z)=\binom{v_X}{v_Z}$, with matrices acting from the left. We occasionally write $(v_X,v_Z)^{\mathsf t}$ to emphasize this.

\subsubsection*{Acknowledgements}
I thank Ahsan Khan for his friendship and encouragement. I appreciate the helpful feedbacks from Nikita Sopenko. I thank Dan Freed and the Center of Mathematical Sciences and Applications at Harvard University for their support and the stimulating environment. My research is supported by an AMS--Simons Travel Grant and by the Simons Collaboration on Global Categorical Symmetries.

\subsubsection*{Use of AI tools}
The invariant $I$ is due to the author. The computations in Section~\ref{sec:examples} were carried out with the assistance of Claude Fable~5.1 and Claude Opus~5.5 (Anthropic), which were also used in preparing the text. ChatGPT Astra (OpenAI) was used to check the arguments. The author takes full responsibility for the content of this paper.

\section{Spin systems, QCA, circuits, and the GNVW index}\label{sec:prelim}

\subsection{Spin systems and QCA}
Throughout, $X$ is a countable set equipped with a metric $d$ in which every ball is finite; the examples are $X=\Z^n$ with the $\ell^\infty$-metric, the cylinders $\Z\times\Z/M$ with the quotient metric, and subsets of these with the induced metric. For $Y\subseteq X$ and $r\ge0$ write $D_r(Y)=\{x\in X:d(x,Y)\le r\}$. A \emph{spin system} on $X$ is a function $p\colon X\to\Z_{\ge1}$; its \emph{local operator algebra} is the algebraic tensor product
\[
\cA(X,p)=\bigotimes_{x\in X}\Mat_{p_x}(\C),
\]
i.e.\ the unital $*$-algebra spanned by operators acting nontrivially on finitely many sites. For $Y\subseteq X$ we regard $\cA(Y,p)$ as the subalgebra of operators supported in $Y$, and $\supp(a)$ denotes the smallest finite set $Y$ with $a\in\cA(Y,p)$. For spin systems $p,q$ on $X$ we identify $\cA(X,p)\otimes\cA(X,q)=\cA(X,pq)$ site by site, where $pq$ is the pointwise product; in particular $\cA(X,p)^{\otimes n}=\cA(X,p^n)$, and we write $a^{(s)}$ for the operator $a\in\cA(X,p)$ placed in the $s$-th tensor factor.

A $*$-homomorphism $\phi$ between subalgebras of local operator algebras on $X$ is \emph{locality preserving with spread $\ell$} if $\supp\phi(a)\subseteq D_\ell(\supp a)$ for all $a$. A \emph{QCA} of $\cA(X,p)$ is a locality preserving $*$-automorphism $\alpha$; the smallest admissible $\ell$ is the \emph{spread} of $\alpha$. The inverse of a QCA of spread $\ell$ has spread $\le\ell$ (cf.\ \cite[Lemma~3.2]{Haah-ISA}): if $a$ is supported in $Y$ and $b$ in $X\setminus D_\ell(Y)$, then $\alpha(b)$ is supported outside $Y$ and commutes with $a$, so $\alpha^{-1}(a)$ commutes with every operator supported in $X\setminus D_\ell(Y)$ and is therefore supported in $D_\ell(Y)$. The QCA of $\cA(X,p)$ form a group $\QCA(X,p)$.

A \emph{layer} of block size $b$ is an automorphism of the form $\prod_B\alpha_B$, where $X=\bigsqcup B$ is a partition into finite sets of diameter $\le b$ and $\alpha_B$ is an automorphism of $\cA(B,p)$ extended by the identity. A \emph{circuit} of depth $m$ and block size $b$ is a composition of $m$ layers of block size $\le b$. Circuits are QCA, and they form a subgroup $\Circ(X,p)\subseteq\QCA(X,p)$. For $X=\Z^n$, $v\in\Z^n$ and a spin system $p$ with $p_{x+v}=p_x$, the \emph{translation} $T_v$ is the QCA of $\cA(\Z^n,p)$ that moves every operator by $v$. Finally, for a QCA $\alpha$ of $\cA(X,p)$ and a spin system $q$ on $X$, the \emph{stabilization} $\alpha\otimes\id_{\cA(X,q)}$ is a QCA of $\cA(X,pq)$.

\begin{definition}\label{def:stably-circuit}
A QCA $\alpha$ of $\cA(\Z^n,p)$ is \emph{stably a circuit} if $\alpha\otimes\id_{\cA(\Z^n,q)}\in\Circ(\Z^n,pq)$ for some spin system $q$ on $\Z^n$, and \emph{stably a circuit modulo translations} if $\alpha\otimes\id_{\cA(\Z^n,q)}=c\circ T_v$ for some $q$, some circuit $c$ and some translation $T_v$ of $\cA(\Z^n,pq)$.
\end{definition}

\subsection{Chains and the GNVW index}\label{subsec:gnvw}
A \emph{chain} is a spin system on $X=\Z$ (with the standard metric); the cells $\cA(\{i\},p)=\Mat_{p_i}(\C)$ may have different dimensions. For a unital $*$-subalgebra $\cB\subseteq\cA_1\otimes\cA_2$ of a tensor product of finite-dimensional matrix algebras, the \emph{support algebra} $\cS(\cB,\cA_1)$ is the smallest unital $*$-subalgebra $\cS\subseteq\cA_1$ with $\cB\subseteq\cS\otimes\cA_2$; it is generated, as a unital $*$-algebra, by the operators $(\id\otimes f)(b)$ with $b\in\cB$ and $f\in\cA_2^*$ a linear functional \cite[Lemma~7]{GNVW}.

Let $\alpha$ be a QCA of a chain $\cA(\Z,p)$ with spread $\ell$, let $n\in\Z$ and let $K\ge\max(\ell,1)$ be an integer. Then $\alpha$ maps $\cA([n-K,n+K-1],p)$ into $\cA([n-2K,n+2K-1],p)=\cA([n-2K,n-1],p)\otimes\cA([n,n+2K-1],p)$, and we consider the support algebra
\[
\mathcal R:=\cS\Big(\alpha\big(\cA([n-K,n+K-1],p)\big),\ \cA([n,n+2K-1],p)\Big)\subseteq\cA([n,n+2K-1],p)
\]
of the image of the two blocks $[n-K,n-1]$ and $[n,n+K-1]$ on the two blocks $[n,n+K-1]$ and $[n+K,n+2K-1]$. By \cite[\S7.2]{GNVW}, $\mathcal R$ is a full matrix algebra, $\mathcal R\cong\Mat_r(\C)$, and we define
\begin{equation}\label{eq:gnvw}
\ind(\alpha):=\frac{r}{p_n\,p_{n+1}\cdots p_{n+K-1}}\ ,
\end{equation}
the ratio of the matrix size of $\mathcal R$ to that of $\cA([n,n+K-1],p)$. Grouping the chain into blocks of $K$ consecutive cells, two of which are $[n-K,n-1]$ and $[n,n+K-1]$, makes $\alpha$ a nearest-neighbour QCA of the block chain; $\mathcal R$ is the algebra $R_{2x+1}$ of \cite[eq.~(24)]{GNVW} for this pair of blocks, and \eqref{eq:gnvw} is the index of \cite[eq.~(28)]{GNVW}, up to replacing it by its inverse, which depends only on the orientation convention. We orient it so that translations to the right have index $>1$ (part (d) below). The following is the main theorem of \cite{GNVW}.

\begin{theorem}[Gross--Nesme--Vogts--Werner]\label{thm:gnvw}
Let $\alpha,\beta$ be QCA of a chain $\cA(\Z,p)$.
\begin{enumerate}[label=\textup{(\alph*)}]
\item $\ind(\alpha)\in\Qpos$ is independent of $n$ and of $K\ge\max(\ell,1)$.
\item $\ind(\alpha\beta)=\ind(\alpha)\ind(\beta)$; in particular $\ind(\beta\alpha\beta^{-1})=\ind(\alpha)$.
\item $\ind(\alpha)=1$ if $\alpha$ is a layer; hence $\ind$ is trivial on circuits.
\item If $p$ is constant, $\ind(T_1)=p$; more generally, if $\cA(\Z,p)=\cA(\Z,p')\otimes\cA(\Z,p'')$ with $p'$ constant and $\alpha=T_1\otimes\id$, then $\ind(\alpha)=p'$.
\item If $\alpha'$ is a QCA of a chain $\cA(\Z,p')$, then $\alpha\otimes\alpha'$ is a QCA of $\cA(\Z,pp')$ and $\ind(\alpha\otimes\alpha')=\ind(\alpha)\ind(\alpha')$; in particular $\ind(\alpha\otimes\id)=\ind(\alpha)$.
\end{enumerate}
\end{theorem}

\subsection{QCA supported near a line}\label{subsec:near-line}
Let $q$ be a spin system on $\Z^2$ and $L=\{x_2=0\}$. For $R\ge0$ let $S_R=\Z\times[-R,R]$ be the strip of half-width $R$ around $L$, and let $\QCA_R(L)$ be the set of QCA $\gamma$ of $\cA(\Z^2,q)$ such that $\gamma(a)=a$ for every $a$ supported in $\Z^2\setminus S_R$. Such a $\gamma$ preserves the commutant $\cA(S_R,q)$ of $\cA(\Z^2\setminus S_R,q)$, and its restriction to $\cA(S_R,q)$ is a QCA of the chain whose $i$-th cell is $\cA(\{i\}\times[-R,R],q)$. We define $\ind(\gamma)$ to be the GNVW index of this chain QCA. By Theorem~\ref{thm:gnvw}(e) the value does not change if $R$ is enlarged ($\gamma$ then becomes the tensor product with the identity on the additional rows), so $\ind$ is well defined on the group
\[
\QCA(L):=\bigcup_{R\ge0}\QCA_R(L)
\]
of QCA supported near $L$, and by Theorem~\ref{thm:gnvw}(b) it is a homomorphism $\ind\colon\QCA(L)\to\Qpos$, trivial on circuits supported near $L$ and invariant under conjugation by elements of $\QCA(L)$.

We shall also use cylinders. For $M\ge1$ let $X_M=\Z\times\Z/M$, with the quotient metric, and let $q_M$ be any spin system on $X_M$ that agrees with $q$ on $\Z\times[-R',R']$ for some $R'$ with $2R'+1<M$ (we identify $[-R',R']$ with its image in $\Z/M$). A QCA $\Gamma$ of $\cA(X_M,q_M)$ is at the same time a QCA of the chain whose $i$-th cell is $\cA(\{i\}\times\Z/M,q_M)$, and $\ind(\Gamma)$ denotes its GNVW index. For $\gamma\in\QCA_R(L)$ with $R\le R'$, let $\gamma_M$ be the QCA of $\cA(X_M,q_M)$ that acts as $\gamma$ on $\cA(S_R,q)=\cA(\Z\times[-R,R],q_M)$ and as the identity on the remaining rows; then $\ind(\gamma_M)=\ind(\gamma)$ by Theorem~\ref{thm:gnvw}(e).

\section{Invertible subalgebras, boundary algebras and pumping}\label{sec:isa}

In this section we recall the parts of Haah's theory \cite{Haah-ISA} that we use and derive the reduction of Theorem A to a statement about the Brauer group. Throughout, $\cA=\cA(\Z^d,p)$ for a spin system $p$ on $\Z^d$.

\subsection{Invertible subalgebras and the Brauer group}
\begin{definition}[{\cite[Def.~1.1]{Haah-ISA}}]\label{def:isa}
A unital $*$-subalgebra $\cB\subseteq\cA$ with commutant $\cC\subseteq \cA$ is an \emph{invertible subalgebra} of spread $\ell$ if every $a\in\cA$ can be written as a finite sum $a=\sum_ib_ic_i$ with $b_i\in\cB$, $c_i\in\cC$ and $\supp b_i,\supp c_i\subseteq D_\ell(\supp a)$.
\end{definition}

\begin{theorem}[{\cite[Lemma~2.4, Cor.~2.5, Prop.~2.7]{Haah-ISA}}]\label{thm:haah-basic}
Let $\cB\subseteq\cA$ be an invertible subalgebra of spread $\ell$ with commutant $\cC$. Then the multiplication map $\mul\colon\cB\otimes\cC\to\cA$ is a $*$-isomorphism, the commutant $\cC$ is again an invertible subalgebra of spread $\ell$.
\end{theorem}

The basic examples are the \emph{trivial} invertible subalgebras: $\cA$ itself (with commutant $\C1$), and more generally a tensor factor $\cA(\Z^d,p_1)\otimes1\subseteq\cA(\Z^d,p_1p_2)$, with commutant $1\otimes\cA(\Z^d,p_2)$. If $\cB\subseteq\cA(\Z^d,p)$ and $\cB'\subseteq\cA(\Z^d,p')$ are invertible with commutants $\cC,\cC'$, then $\cB\otimes\cB'\subseteq\cA(\Z^d,pp')$ is invertible with commutant $\cC\otimes\cC'$. In particular, for any spin system $q$, the subalgebra $\cB\otimes\cA(\Z^d,q)\subseteq\cA(\Z^d,pq)$ is invertible with commutant $\cC\otimes1$; we call it the \emph{stabilization} of $\cB$ by $q$.

\begin{definition}[{\cite[Def.~2.9]{Haah-ISA}}]\label{def:stable-equiv}
Invertible subalgebras $\cB_1\subseteq\cA(\Z^d,p_1)$ and $\cB_2\subseteq\cA(\Z^d,p_2)$ are \emph{stably equivalent} if there are spin systems $q_1,q_2$ on $\Z^d$ and a $*$-isomorphism
\[
\phi\colon\cB_1\otimes\cA(\Z^d,q_1)\longrightarrow\cB_2\otimes\cA(\Z^d,q_2)
\]
which is locality preserving together with its inverse. The set $\Br_d$ of stable equivalence classes is an abelian group under tensor product \cite[Prop.~2.10]{Haah-ISA}, the \emph{Brauer group}; its zero element is the class of the trivial invertible subalgebras, and $[\cC]=-[\cB]$ for the commutant $\cC$ of $\cB$.
\end{definition}

In particular $[\cB]=0$ if and only if there are spin systems $q_1,q_2$ and a locality-preserving isomorphism $\cB\otimes\cA(\Z^d,q_1)\cong\cA(\Z^d,q_2)$ with locality-preserving inverse.

\subsection{Boundary algebras and pumping QCA}
Now let $d\ge1$, write $\Z^{d+1}=\Z\times\Z^d$ with coordinates $(x_1,y)$, and let $\alpha$ be a QCA of $\cA(\Z^{d+1},q)$ of spread $\ell$. For $n\in\Z$ let $H_n=\{x_1\le n\}$ be the half-space and $S_n=\{n-\ell\le x_1\le n+\ell\}$ the slab around its boundary. The \emph{boundary algebra} of $\alpha$ at $n$ is
\[
\cB(\alpha,n):=\alpha\big(\cA(H_n,q)\big)\cap\cA(S_n,q).
\]
We regard $\cA(S_n,q)$ as the local operator algebra $\cA(\Z^d,q_n)$ of the spin system $q_n(y)=\prod_{x_1=n-\ell}^{n+\ell}q(x_1,y)$ on $\Z^d$.

\begin{theorem}[{\cite[Lemmas~3.3, 3.6, 3.7]{Haah-ISA}}]\label{thm:haah-boundary}
$\cB(\alpha,n)$ is an invertible subalgebra of $\cA(\Z^d,q_n)$, and $\alpha(\cA(H_n,q))=\cA(H_{n-\ell-1},q)\otimes\cB(\alpha,n)$, where $H_{n-\ell-1}=\{x_1\le n-\ell-1\}$ is disjoint from $S_n$ and $H_{n+\ell}=H_{n-\ell-1}\sqcup S_n$. Its stable equivalence class
\[
\partial\alpha:=[\cB(\alpha,n)]\in\Br_d
\]
is independent of $n$ and of the choice of $\ell\ge\mathrm{spread}(\alpha)$.
\end{theorem}

\begin{lemma}\label{lem:boundary-local}
Let $\alpha,\beta$ be QCA of $\cA(\Z^{d+1},q)$ and $n_0\in\Z$. If $\alpha(a)=\beta(a)$ for every $a$ supported in $\{x_1>n_0\}$, or for every $a$ supported in $\{x_1<n_0\}$, then $\partial\alpha=\partial\beta$.
\end{lemma}
\begin{proof}
Let $\ell$ bound the spreads of $\alpha$ and $\beta$ and suppose first that they agree on $\{x_1>n_0\}$. Take $n>n_0+2\ell$. For $a\in\cA(S_n,q)$, the operator $b:=\alpha^{-1}(a)$ is supported in $D_\ell(S_n)\subseteq\{x_1>n_0\}$, so $\beta(b)=\alpha(b)=a$ and $\beta^{-1}(a)=\alpha^{-1}(a)$. Hence $a\in\cB(\alpha,n)$ if and only if $\alpha^{-1}(a)\in\cA(H_n,q)$, if and only if $\beta^{-1}(a)\in\cA(H_n,q)$, if and only if $a\in\cB(\beta,n)$. So $\cB(\alpha,n)=\cB(\beta,n)$ and $\partial\alpha=\partial\beta$ by Theorem~\ref{thm:haah-boundary}. If instead $\alpha,\beta$ agree on $\{x_1<n_0\}$, take $n<n_0-2\ell$: then $\alpha$ and $\beta$ agree on $\cA(H_n,q)$, so $\cB(\alpha,n)=\cB(\beta,n)$ directly.
\end{proof}

\begin{lemma}\label{lem:boundary-trivial}
Let $\alpha$ be a QCA of $\cA(\Z^{d+1},q)$.
\begin{enumerate}[label=\textup{(\roman*)}]
\item $\partial(\alpha\otimes\id_{\cA(\Z^{d+1},q')})=\partial\alpha$ for every spin system $q'$ on $\Z^{d+1}$.
\item $\partial T_v=0$ for every translation $T_v$ of $\cA(\Z^{d+1},q)$.
\item $\partial(c\circ\alpha)=\partial\alpha$ for every circuit $c$ of $\cA(\Z^{d+1},q)$.
\end{enumerate}
Consequently $\partial(c\circ T_v)=0$ for every circuit $c$ and translation $T_v$.
\end{lemma}
\begin{proof}
(i) For subspaces $M,N\subseteq V$ and $M',N'\subseteq V'$ of vector spaces one has $(M\otimes M')\cap(N\otimes N')=(M\cap N)\otimes(M'\cap N')$. Applying this to $\alpha(\cA(H_n,q))\otimes\cA(H_n,q')$ and $\cA(S_n,q)\otimes\cA(S_n,q')$ gives $\cB(\alpha\otimes\id,n)=\cB(\alpha,n)\otimes\cA(H_n\cap S_n,q')$, and $\cA(H_n\cap S_n,q')$ is the full local operator algebra of a spin system on $\Z^d$. Thus $\cB(\alpha\otimes\id,n)$ is a stabilization of $\cB(\alpha,n)$ and has the same class.

(ii) Choose $\ell\ge|v|$. If $v_1\ge0$ then $T_v(\cA(H_n,q))=\cA(H_{n+v_1},q)$ and $\cB(T_v,n)=\cA(H_{n+v_1}\cap S_n,q)$; if $v_1<0$ then $\cB(T_v,n)=\cA(H_{n+v_1}\cap S_n,q)$ as well. In both cases $\cB(T_v,n)$ is a tensor factor of $\cA(S_n,q)$ which is itself a full local operator algebra of a spin system on $\Z^d$, hence trivial: it is stably equivalent (with $q_1=q_2=1$ and $\phi$ the identity map) to the full algebra of that spin system.

(iii) Write $c=c_1\cdots c_m$ with layers $c_j$ of block size $b$, and let $c_j^+$ be the layer obtained from $c_j$ by keeping only the blocks contained in $\{x_1>n_0\}$, $c^+:=c_1^+\cdots c_m^+$. Let $\ell$ be the spread of $\alpha$. If $a$ is supported in $\{x_1>n_0+mb+\ell\}$, then $\alpha(a)$ is supported in $\{x_1>n_0+mb\}$, and inductively $c_m^+$, then $c_{m-1}^+$, \dots\ act on it as $c_m,c_{m-1},\dots$ (at each stage the operator is supported in $\{x_1>n_0+(j-1)b\}$ and meets only blocks contained in $\{x_1>n_0\}$); so $c^+\alpha(a)=c\alpha(a)$. If $a$ is supported in $\{x_1<n_0-\ell\}$, then $\alpha(a)$ is supported in $\{x_1<n_0\}$ and meets no block of any $c_j^+$, so $c^+\alpha(a)=\alpha(a)$. By Lemma~\ref{lem:boundary-local}, applied once on the right and once on the left, $\partial(c\alpha)=\partial(c^+\alpha)=\partial\alpha$.
\end{proof}

\begin{remark}\label{rem:gen-shift}
More generally, let $s$ be a \emph{generalized shift}: an automorphism of $\cA(\Z^{d+1},q)$ that permutes the factors of a fixed on-site tensor decomposition $\cA(\{x\},q)=\bigotimes_j\cA_{x,j}$, moving each factor by a uniformly bounded distance. Then $s(\cA(H_n,q))$ is the tensor product of some of these factors, so $\cB(s,n)$ is the tensor product of those of them that lie in $S_n$, a full local operator algebra of a spin system on $\Z^d$; hence $\partial s=0$. Consequently Proposition~\ref{prop:nontrivial-from-Brauer} and Theorem~\ref{thm:main} also exclude compositions of stabilized circuits with generalized shifts.
\end{remark}

\begin{construction}[Pumping QCA, {\cite[Lemma~3.9]{Haah-ISA}}]\label{constr:pumping}
Let $\cB\subseteq\cA(\Z^d,p)$ be an invertible subalgebra of spread $\ell$ with commutant $\cC$, and let $\hat p$ be the spin system $\hat p(x_1,y)=p(y)$ on $\Z^{d+1}=\Z\times\Z^d$, so that $\cA(\Z^{d+1},\hat p)=\bigotimes_{j\in\Z}\cA^{(j)}$ with $\cA^{(j)}=\cA(\{j\}\times\Z^d,\hat p)\cong\cA(\Z^d,p)$ the $j$-th \emph{sheet}. By Theorem~\ref{thm:haah-basic}, $\cA(\Z^{d+1},\hat p)=\bigotimes_j\cB^{(j)}\otimes\bigotimes_j\cC^{(j)}$. The \emph{pumping QCA} $\alpha_\cB$ is the $*$-automorphism determined by
\[
\alpha_\cB(b^{(j)})=b^{(j+1)}\quad(b\in\cB),\qquad\alpha_\cB(c^{(j)})=c^{(j)}\quad(c\in\cC).
\]
It is a QCA of spread $\ell+1$: an operator $a^{(j)}$ with $a\in\cA(\Z^d,p)$ decomposes as $\sum_ib_i^{(j)}c_i^{(j)}$ with $b_i,c_i$ supported in $D_\ell(\supp a)$, and $\alpha_\cB(a^{(j)})=\sum_ib_i^{(j+1)}c_i^{(j)}$; the inverse moves $\cB$ back by one sheet.
\end{construction}

\begin{proposition}[{\cite[Lemma~3.9]{Haah-ISA}}]\label{prop:pumping-boundary}
$\partial\alpha_\cB=[\cB]$.
\end{proposition}
\begin{proof}
With $\ell$ replaced by $\ell+1$ in the definition of $S_n$, so that $S_n=\{n-\ell-1\le x_1\le n+\ell+1\}$, we have $\alpha_\cB(\cA(H_n,\hat p))=\bigotimes_{j\le n+1}\cB^{(j)}\otimes\bigotimes_{j\le n}\cC^{(j)}=\cA(H_n,\hat p)\otimes\cB^{(n+1)}=\cA(H_{n-\ell-2},\hat p)\otimes\cA(H_n\cap S_n,\hat p)\otimes\cB^{(n+1)}$, so $\cB(\alpha_\cB,n)=\cA(H_n\cap S_n,\hat p)\otimes\cB^{(n+1)}$, which is a stabilization of $\cB$ (by the full local algebra of the sheets $n-\ell-1,\dots,n$), hence stably equivalent to $\cB$.
\end{proof}

\subsection{Reduction of Theorem A}
\begin{proposition}\label{prop:nontrivial-from-Brauer}
Let $\cB\subseteq\cA(\Z^d,p)$ be an invertible subalgebra with $[\cB]\ne0$ in $\Br_d$, $d\ge1$. Then the pumping QCA $\alpha_\cB$ of $\cA(\Z^{d+1},\hat p)$ is not stably a circuit modulo translations. More generally, the same holds for every QCA $\alpha$ of $\cA(\Z^{d+1},q)$ with $\partial\alpha=[\cB]$.
\end{proposition}
\begin{proof}
If $\alpha\otimes\id_{q'}=c\circ T_v$, then by Lemma~\ref{lem:boundary-trivial} $[\cB]=\partial\alpha=\partial(\alpha\otimes\id_{q'})=\partial(c\circ T_v)=0$, a contradiction. For $\alpha=\alpha_\cB$ use Proposition~\ref{prop:pumping-boundary}.
\end{proof}

Thus Theorem A follows once we show $[\cB_{p,k}]\ne0$ in $\Br_2$. The rest of the paper is devoted to an invariant of two-dimensional invertible subalgebras capable of showing this.

\section{The half-plane invariant}\label{sec:invariant}

Throughout this section $\cA=\cA(\Z^2,p)$ and $\cB\subseteq\cA$ is an invertible subalgebra of spread $\ell$ with commutant $\cC$. We write $L=\{x_2=0\}$ and $P=\{x_2\ge0\}$ for the line and the upper half-plane.

\subsection{The Brauer swap}
 $\cA^{\otimes2}=\cA(\Z^2,p^2)$ is the tensor product $\cB^{(1)}\otimes\cB^{(2)}\otimes\cC^{(1)}\otimes\cC^{(2)}$ of the two copies of $\cB$ and the two copies of $\cC$.

\begin{definition}\label{def:brauer-swap}
The \emph{Brauer swap} of $\cB$ is the $*$-automorphism $\tau_\cB$ of $\cA^{\otimes2}$ determined by
\[
\tau_\cB(b^{(1)})=b^{(2)},\qquad\tau_\cB(b^{(2)})=b^{(1)},\qquad\tau_\cB(c^{(s)})=c^{(s)}\qquad(b\in\cB,\ c\in\cC,\ s=1,2).
\]
\end{definition}
\

\begin{lemma}\label{lem:swap-basic}
\begin{enumerate}[label=\textup{(\roman*)}]
\item $\tau_\cB$ is an involutive QCA of spread $\ell$.
\item If $\cB=\cA$, then $\tau_\cB$ is the on-site swap $S$ of the two tensor factors; if $\cB=\cA(\Z^2,p_1)\otimes1\subseteq\cA(\Z^2,p_1p_2)$, then $\tau_\cB$ is the on-site swap of the two copies of the first factor.
\item For a spin system $q$, the Brauer swap of the stabilization $\cB\otimes\cA(\Z^2,q)\subseteq\cA(\Z^2,pq)$ is $\tau_\cB\otimes S_q$, where $S_q$ is the on-site swap of the two copies of $\cA(\Z^2,q)$.
\item For invertible subalgebras $\cB\subseteq\cA(\Z^2,p)$ and $\cB'\subseteq\cA(\Z^2,p')$, the Brauer swap of $\cB\otimes\cB'\subseteq\cA(\Z^2,pp')$ is $\sigma\,(\tau_\cB\otimes\tau_{\cB'})\,\sigma^{-1}$, where $\sigma$ is the on-site permutation of tensor factors identifying $\cA(\Z^2,p)^{\otimes2}\otimes\cA(\Z^2,p')^{\otimes2}$ with $\cA(\Z^2,pp')^{\otimes2}$.
\item If $\cB$ is invariant under a translation $T_v$ of $\cA$, then $\tau_\cB$ commutes with $T_v\otimes T_v$.
\end{enumerate}
\end{lemma}
\begin{proof}
(i) $\tau_\cB$ is well defined and involutive as the automorphism of a tensor product of four factors that exchanges two of them. For $a\in\cA^{\otimes2}$ supported in a finite set $Y$, the decomposition of Theorem~\ref{thm:haah-basic} writes $a$ as a sum of products $b_1^{(1)}b_2^{(2)}c_1^{(1)}c_2^{(2)}$ with all factors supported in $D_\ell(Y)$, and $\tau_\cB$ permutes these factors, so $\supp\tau_\cB(a)\subseteq D_\ell(Y)$. (ii)--(v) are immediate from the definition.
\end{proof}

\subsection{Truncations}
\begin{definition}\label{def:truncation}
Let $\tau$ be an involutive QCA of $\cA(\Z^2,q)$ and $m\in\Z$. A \emph{truncation of $\tau$ to $\{x_2\ge m\}$} is a QCA $t$ of $\cA(\Z^2,q)$ for which there is an $r\ge0$ such that
\[
t(a)=\tau(a)\ \text{ for $a$ supported in }\{x_2>m+r\},\qquad t(a)=a\ \text{ for $a$ supported in }\{x_2<m-r\}.
\]
We call $\tau$ \emph{truncatable} if there are $r_0,s_0\ge0$ such that for every $m\in\Z$ it admits a truncation to $\{x_2\ge m\}$ with parameter $r\le r_0$ and spread $\le s_0$. 
\end{definition}

If $t$ is a truncation of $\tau$ to $\{x_2\ge0\}$ with parameter $r$, and $s$ bounds the spreads of $t$ and $\tau$, then for $a$ supported in $\{x_2>r+s\}$ we have $t^2(a)=t(\tau(a))=\tau^2(a)=a$, and for $a$ supported in $\{x_2<-r\}$ we have $t^2(a)=a$. Hence
\[
t^2\in\QCA_{r+s}(L)\qquad\text{and}\qquad\ind(t^2)\in\Qpos.
\]

\begin{lemma}[Conjugation by a truncation]\label{lem:conj-truncation}
Let $\tau$ be a truncatable involutive QCA of $\cA(\Z^2,q)$, let $t$ be a truncation of $\tau$ to $\{x_2\ge0\}$, and let $\gamma\in\QCA(L)$. Then $t\gamma t^{-1}\in\QCA(L)$ and $\ind(t\gamma t^{-1})=\ind(\gamma)$.
\end{lemma}
\begin{proof}
Let $r$ and $s$ be as above and $\gamma\in\QCA_R(L)$; clearly $t\gamma t^{-1}\in\QCA_{R+2s}(L)$. Let $r_0,s_0$ be as in Definition~\ref{def:truncation}. Choose $M\in8\Z$ large compared with $r,s,R,r_0,s_0$, put $m:=M/2$, and choose a truncation $t'$ of $\tau$ to $\{x_2\ge m\}$ with parameter $r'\le r_0$ and spread $s'\le s_0$; thus $r',s'$ are small compared with $M$. The QCA $u:=\tau t'^{-1}$ satisfies $u(a)=a$ for $a$ supported in $\{x_2>m+r'+s'\}$ and $u(a)=\tau(a)$ for $a$ supported in $\{x_2<m-r'-s'\}$.

Consider the cylinder $X_M=\Z\times\Z/M$, identify $\Z/M$ with the rows $-M/4,\dots,3M/4-1$ of $\Z^2$, and let $q_M$ be the restriction of $q$ to these rows. Cover the cylinder by the two bands $U_1=\Z\times(-3M/8,3M/8)$ and $U_2=\Z\times(M/8,7M/8)$, which overlap in the rows $(M/8,3M/8)$ and $(5M/8,7M/8)\equiv(-3M/8,-M/8)$. On operators supported in $U_1$ the QCA $t$ is defined (we regard $U_1$ as a subset of $\Z^2$), on operators supported in $U_2$ the QCA $u$ is defined, and on operators supported in the overlaps both agree. Define $\Gamma$ on the matrix algebra of each site of the cylinder by $t$ or $u$, whichever applies, and extend multiplicatively. The relations among single-site operators are the matrix relations at each site and the commutation of operators at distinct sites; the former are preserved because $\Gamma$ is a $*$-homomorphism on each site algebra, and the latter because two sites at row distance $\le M/8$ lie in a common band, where $\Gamma$ is given by a single automorphism, while for two sites at row distance $>M/8\ge2\max(s,s')$ the images have disjoint supports. Hence $\Gamma$ is a QCA of $\cA(X_M,q_M)$. Since $\Gamma$ agrees with $t$ on all operators supported in $U_1$, and $\Gamma^{-1}$ with $t^{-1}$ on operators supported in the rows $(-3M/8+s,3M/8-s)$, the QCA $\Gamma\gamma_M\Gamma^{-1}$ and $(t\gamma t^{-1})_M$ agree on all operators supported in $S_{R+2s}$, and both are the identity on operators supported outside $S_{R+2s}$. Hence $\ind(t\gamma t^{-1})=\ind\big((t\gamma t^{-1})_M\big)=\ind(\Gamma\gamma_M\Gamma^{-1})=\ind(\gamma_M)=\ind(\gamma)$.
\end{proof}

\begin{proposition}[Independence of the truncation]\label{prop:well-defined}
Let $\tau$ be a truncatable involutive QCA of $\cA(\Z^2,q)$ and let $t_1,t_2$ be truncations of $\tau$ to $\{x_2\ge0\}$. Then $\ind(t_2^2)\,\ind(t_1^2)^{-1}\in\Qpos^2$.
\end{proposition}
\begin{proof}
Put $\gamma:=t_1^{-1}t_2$. Since $t_1$ and $t_2$ both agree with $\tau$ far above $L$ and with the identity far below $L$, $\gamma\in\QCA(L)$. Then
\[
t_2^2=t_1\gamma t_1\gamma=t_1^2\cdot(t_1^{-1}\gamma t_1)\cdot\gamma,
\]
and all three factors on the right lie in $\QCA(L)$. By the homomorphism property of $\ind$ on $\QCA(L)$ and Lemma~\ref{lem:conj-truncation}, $\ind(t_2^2)=\ind(t_1^2)\ind(\gamma)^2$.
\end{proof}

\begin{definition}\label{def:invariant}
For a truncatable involutive QCA $\tau$ of $\cA(\Z^2,q)$ put
\[
I(\tau):=\big[\ind(t^2)\big]\in\Qpos/\Qpos^2,
\]
where $t$ is any truncation of $\tau$ to $\{x_2\ge0\}$. For an invertible subalgebra $\cB\subseteq\cA(\Z^2,p)$ whose Brauer swap is truncatable we put $I(\cB):=I(\tau_\cB)$.
\end{definition}

\begin{remark}\label{rem:ti-truncatable}
If $\cB$ is invariant under the translations of $\Z^2$ (or merely under a translation $T_v$ with $v_2\ne0$), then $\tau_\cB$ is truncatable (with uniform bounds) as soon as it admits a single truncation $t$, say to $\{x_2\ge0\}$ with parameter $r$. Indeed, by Lemma~\ref{lem:swap-basic}(v), $(T_v\otimes T_v)^jt(T_v\otimes T_v)^{-j}$ is a truncation of $\tau_\cB$ to $\{x_2\ge jv_2\}$ with parameter $r$ and the same spread, and a truncation to $\{x_2\ge m\}$ with parameter $r$ is a truncation to $\{x_2\ge m'\}$ with parameter $r+|m-m'|$.
\end{remark}

\begin{remark}[Existence of truncations in general]\label{rem:existence}
Truncations exist for every invertible subalgebra after a suitable stabilization, but the only proof we know uses the classification of two-dimensional QCA~ \cite{FH}. We do not need any of this as truncations of $\tau_\cB$ are constructed explicitly in Section~\ref{sec:examples}. For the remainder of the article, we retain the assumption that the Brauer swap is truncatable, to avoid invoking the general existence result.
\end{remark}

\subsection{Properties of the invariant}\label{subsec:properties}
\begin{lemma}\label{lem:I-stabilization}
Let $\cB\subseteq\cA(\Z^2,p)$ be an invertible subalgebra whose Brauer swap is truncatable, and let $q$ be a spin system on $\Z^2$.
\begin{enumerate}[label=\textup{(\roman*)}]
\item The Brauer swap $\tau_\cB\otimes S_q$ of the stabilization $\cB\otimes\cA(\Z^2,q)$ is truncatable, and $I(\cB\otimes\cA(\Z^2,q))=I(\cB)$.
\item For every on-site involution $\psi=\prod_x\Ad(w_x)$ of $\cA(\Z^2,q)$, $w_x\in\cA(\{x\},q)$, $w_x^2=1$ (including $\psi=\id$), the involutive QCA $\tau_\cB\otimes\psi$ of $\cA(\Z^2,p^2q)$ is truncatable, and $I(\tau_\cB\otimes\psi)=I(\cB)$.
\item The Brauer swap of the commutant $\cC$ is truncatable.
\end{enumerate}
\end{lemma}
\begin{proof}
(i) If $t$ is a truncation of $\tau_\cB$ to $\{x_2\ge m\}$, then $t\otimes S_{q,\ge m}$, where $S_{q,\ge m}$ is the product of the on-site swaps at the sites of $\{x_2\ge m\}$, is a truncation of $\tau_\cB\otimes S_q$ to $\{x_2\ge m\}$, and $(t\otimes S_{q,\ge0})^2=t^2\otimes\id$ has the same index as $t^2$ by Theorem~\ref{thm:gnvw}(e). (ii) Similarly with $t\otimes\psi_{\ge m}$, $\psi_{\ge m}:=\prod_{x_2\ge m}\Ad(w_x)$: this is a truncation of $\tau_\cB\otimes\psi$, and $(t\otimes\psi_{\ge0})^2=t^2\otimes\id$. (iii) Since $\tau_\cB\tau_\cC=S$ is the on-site swap of the two copies of $\cA$, $\tau_\cC=\tau_\cB S$. If $t$ is a truncation of $\tau_\cB$ to $\{x_2\ge m\}$, then $tS_{\ge m}$ is a truncation of $\tau_\cC$ to $\{x_2\ge m\}$.
\end{proof}

\begin{proposition}[Additivity]\label{prop:additive}
Let $\cB\subseteq\cA(\Z^2,p)$ and $\cB'\subseteq\cA(\Z^2,p')$ be invertible subalgebras with truncatable Brauer swaps. Then the Brauer swap of $\cB\otimes\cB'$ is truncatable and
\[
I(\cB\otimes\cB')=I(\cB)+I(\cB')\qquad\text{in }\Qpos/\Qpos^2 .
\]
\end{proposition}
\begin{proof}
Let $t,t'$ be truncations of $\tau_\cB,\tau_{\cB'}$ to $\{x_2\ge m\}$ and let $\sigma$ be the on-site permutation of tensor factors of Lemma~\ref{lem:swap-basic}(iv). Then $\sigma(t\otimes t')\sigma^{-1}$ is a truncation of $\tau_{\cB\otimes\cB'}=\sigma(\tau_\cB\otimes\tau_{\cB'})\sigma^{-1}$ to $\{x_2\ge m\}$, and its square is $\sigma(t^2\otimes t'^2)\sigma^{-1}$. Since $\sigma$ is a layer, in particular an element of $\QCA(L)$ when restricted to any strip, $\ind(\sigma(t^2\otimes t'^2)\sigma^{-1})=\ind(t^2\otimes t'^2)=\ind(t^2)\ind(t'^2)$ by Theorem~\ref{thm:gnvw}(b),(e).
\end{proof}

\begin{proposition}[Vanishing on stably trivial subalgebras]\label{prop:Brauer-invariance}
Let $\cB\subseteq\cA(\Z^2,p)$ be an invertible subalgebra whose Brauer swap is truncatable. If $[\cB]=0$ in $\Br_2$, then $I(\cB)=0$.
\end{proposition}
\begin{proof}
By Definition~\ref{def:stable-equiv} there are spin systems $q_1,q_2$ and a locality-preserving isomorphism with locality-preserving inverse
\[
\phi\colon\cD:=\cB\otimes\cA(\Z^2,q_1)\longrightarrow\cA(\Z^2,q_2),
\]
say of spread $\ell_\phi$ in both directions. Put $\cE:=\cA(\Z^2,pq_1)\supseteq\cD$. The commutant of $\cD$ in $\cE$ is $\cC\otimes1$, and $\cE=\cD\otimes\cC$. For $x\in\Z^2$ let
\[
\cD_x:=\phi^{-1}\big(\Mat_{q_2(x)}(\C)\big)\subseteq\cD
\]
be the preimage of the matrix algebra at the site $x$ of $\cA(\Z^2,q_2)$. The $\cD_x$ are full matrix algebras, supported in $D_{\ell_\phi}(\{x\})$, they commute with each other, and the subalgebra they generate is $\phi^{-1}(\cA(\Z^2,q_2))=\cD$; hence $\cD=\bigotimes_x\cD_x$ (algebraic tensor product), and
\[
\cE^{\otimes2}=\bigotimes_x\big(\cD_x^{(1)}\otimes\cD_x^{(2)}\big)\otimes\cC^{(1)}\otimes\cC^{(2)}.
\]
Let $t$ be the $*$-automorphism of $\cE^{\otimes2}$ that exchanges the two factors $\cD_x^{(1)}$ and $\cD_x^{(2)}$ for every $x\in P$ and is the identity on $\cD_x^{(1)}\otimes\cD_x^{(2)}$ for $x\notin P$ and on $\cC^{(1)}\otimes\cC^{(2)}$. Then $t^2=\id$. We claim that $t$ is a truncation of $\tau_\cD=\tau_\cB\otimes S_{q_1}$ to $\{x_2\ge0\}$. Let $a\in\cE^{\otimes2}$ be supported in a finite set $Y$. By Theorem~\ref{thm:haah-basic}, $a$ is a sum of products $d_1^{(1)}d_2^{(2)}c_1^{(1)}c_2^{(2)}$ with $d_i\in\cD$, $c_i\in\cC$ supported in $D_\ell(Y)$, and since $\phi(d_i)$ is supported in $D_{\ell+\ell_\phi}(Y)$, each $d_i$ is a finite sum of products of elements of the $\cD_x$ with $x\in D_{\ell+\ell_\phi}(Y)$. It follows that $t$ is locality preserving with spread $\ell+2\ell_\phi$, hence (being an involution) a QCA. Clearly, $t(a)=a$ if $Y\subseteq\{x_2<-\ell-\ell_\phi\}$ and $t(a)=\tau_\cD(a)$ if $Y\subseteq\{x_2>\ell+\ell_\phi\}$. By Lemma~\ref{lem:I-stabilization}(i), $\tau_\cD$ is truncatable and $I(\cB)=I(\cD)$. Therefore, $I(\cD)=[\ind(t^2)]=[\ind(\id)]=0$.
\end{proof}

\begin{theorem}[Brauer invariance]\label{thm:Brauer-invariance}
Let $\cB_1\subseteq\cA(\Z^2,p_1)$ and $\cB_2\subseteq\cA(\Z^2,p_2)$ be invertible subalgebras with truncatable Brauer swaps. If $[\cB_1]=[\cB_2]$ in $\Br_2$, then $I(\cB_1)=I(\cB_2)$. Consequently $I$ induces a homomorphism from the subgroup of $\Br_2$ of classes represented by invertible subalgebras with truncatable Brauer swap to $\Qpos/\Qpos^2$.
\end{theorem}
\begin{proof}
Let $\cC_2$ be the commutant of $\cB_2$. Since $[\cB_2\otimes\cC_2]=0$, Propositions~\ref{prop:additive} and \ref{prop:Brauer-invariance} give $I(\cB_2)+I(\cC_2)=0$. Likewise $[\cB_1\otimes\cC_2]=[\cB_1]-[\cB_2]=0$ gives $I(\cB_1)+I(\cC_2)=0$. Hence $I(\cB_1)=-I(\cC_2)=I(\cB_2)$.
\end{proof}

\begin{corollary}[Stable linearizability]\label{cor:onsite}
Let $\cB\subseteq\cA(\Z^2,p)$ be an invertible subalgebra with truncatable Brauer swap. Suppose that for some spin system $q$ on $\Z^2$ there are an on-site involution $\psi=\prod_x\Ad(w_x)$ of $\cA(\Z^2,q)$, a QCA $\beta$ of $\cA(\Z^2,p^2q)$ and an on-site involution $\sigma=\prod_x\Ad(u_x)$ of $\cA(\Z^2,p^2q)$, with $w_x^2=u_x^2=1$ and
\[
\tau_\cB\otimes\psi=\beta\sigma\beta^{-1}.
\]
Then $I(\cB)=0$. In particular, if $I(\cB)\ne0$, then the $\Z/2$-action $\tau_\cB$ on two copies of $\cA(\Z^2,p)$ is neither stably nor weakly linearizable in the sense of \cite{JY-obstruction}.
\end{corollary}
\begin{proof}
Let $\sigma_P:=\prod_{x\in P}\Ad(u_x)$, a layer and an involution. Then $t:=\beta\sigma_P\beta^{-1}$ is a QCA which agrees with $\beta\sigma\beta^{-1}=\tau_\cB\otimes\psi$ on operators supported in $\{x_2>2\,\mathrm{spread}(\beta)\}$ and with the identity on operators supported in $\{x_2<-2\,\mathrm{spread}(\beta)\}$. By Lemma~\ref{lem:I-stabilization}(ii) and Proposition~\ref{prop:well-defined}, $I(\cB)=I(\tau_\cB\otimes\psi)=[\ind(t^2)]=0$.
\end{proof}

\section{Clifford quantum cellular automata}\label{sec:pauli}

In this section $p$ is an odd prime and $\omega=e^{2\pi i/p}$. We collect the facts about Clifford QCA that are used in the computation of Section~\ref{sec:examples}.

\subsection{Pauli operators and Clifford QCA}
On a single qudit $\C^p$ let $X,Z$ be the clock and shift operators, $ZX=\omega XZ$, and for $v=(a,b)\in\F_p^2$ put
\[
W(v):=\omega^{\frac12ab}X^aZ^b,
\]
where $\frac12$ denotes the inverse of $2$ in $\F_p$. A direct computation using $Z^bX^{a'}=\omega^{a'b}X^{a'}Z^b$ gives
\begin{equation}\label{eq:weyl}
W(v)W(w)=\omega^{-\frac12\langle v,w\rangle}W(v+w),\qquad W(v)^*=W(-v),
\end{equation}
where $\langle v,w\rangle:=ab'-a'b$ for $v=(a,b)$, $w=(a',b')$.
Now let $X$ be a countable set and let the spin system have $n_x$ qudits of dimension $p$ at each site $x$, i.e.\ $\cA=\cA(X,p^{n})$. The \emph{Pauli module} is the $\F_p$-vector space
\[
\Pau:=\bigoplus_{x\in X}\F_p^{2n_x}
\]
of finitely supported vectors, with the symplectic form $\langle v,w\rangle=\sum_x\langle v_x,w_x\rangle$, which is nondegenerate: $\langle v,\Pau\rangle=0$ implies $v=0$. For $v\in\Pau$ let $W(v)=\bigotimes_{x}W(v_x)$, then \eqref{eq:weyl} holds verbatim. The operators $W(v)$, $v\in\Pau$, form a basis of $\cA$, and $\supp W(v)=\supp v$. We write $v=(v_X,v_Z)$ for the $X$- and $Z$-components.

A linear map $T\colon\Pau\to\Pau$ is \emph{local of spread $s$} if $\supp Tv\subseteq D_s(\supp v)$ for all $v$, and \emph{symplectic} if $\langle Tv,Tw\rangle=\langle v,w\rangle$. By \eqref{eq:weyl}, for a symplectic bijection $T$ the assignment $W(v)\mapsto W(Tv)$ extends to a $*$-automorphism $\Theta_T$ of $\cA$. If $T$ and $T^{-1}$ are local, $\Theta_T$ is a QCA, which we call the \emph{Clifford QCA} with matrix $T$. Clearly $\Theta_{TT'}=\Theta_T\Theta_{T'}$ and $\supp\Theta_T(a)\subseteq D_s(\supp a)$.

\subsection{Pauli subalgebras}
For an $\F_p$-subspace $G\subseteq\Pau$ let $\cB_G:=\operatorname{span}\{W(v):v\in G\}$. By \eqref{eq:weyl} it is a unital $*$-subalgebra of $\cA$. Let $G^\perp=\{w\in\Pau:\langle w,G\rangle=0\}$. Expanding an operator in the basis $W(w)$ and using $W(v)W(w)W(v)^{-1}=\omega^{-\langle v,w\rangle}W(w)$, one sees that the commutant of $\cB_G$ in $\cA$ is $\cB_{G^\perp}$.

\begin{lemma}\label{lem:pauli-isa}
Let $G\subseteq\Pau$ be a subspace such that $\Pau=G\oplus G^\perp$, and suppose that the projection $\pi_G\colon\Pau\to G$ along $G^\perp$ is local of spread $\ell$. Then $\cB_G$ is an invertible subalgebra of $\cA$ of spread $\ell$ with commutant $\cB_{G^\perp}$.
\end{lemma}
\begin{proof}
The commutant has just been identified. For $v\in\Pau$ write $v=g+h$ with $g=\pi_G(v)\in G$ and $h=v-g\in G^\perp$. Both are supported in $D_\ell(\supp v)$, and $W(v)=\omega^{\frac12\langle g,h\rangle}W(g)W(h)$ with $W(g)\in\cB_G$ and $W(h)\in\cB_{G^\perp}$. A general $a\in\cA$ is a finite linear combination of such $W(v)$ with $\supp v\subseteq\supp a$.
\end{proof}

We call such a $\cB_G$ a \emph{Pauli invertible subalgebra}. Its Brauer swap is Clifford:

\begin{lemma}\label{lem:swap-clifford}
In the situation of Lemma~\textup{\ref{lem:pauli-isa}}, let $\Pau^{(2)}=\Pau\oplus\Pau$ be the Pauli module of $\cA^{\otimes2}$, with the two summands corresponding to the two copies, and let $T_\tau\colon\Pau^{(2)}\to\Pau^{(2)}$ be the linear map
\[
T_\tau(g_1+h_1,\,g_2+h_2)=(g_2+h_1,\,g_1+h_2)\qquad(g_i\in G,\ h_i\in G^\perp).
\]
Then $T_\tau$ is a symplectic involution, local of spread $\ell$, and $\tau_{\cB_G}=\Theta_{T_\tau}$.
\end{lemma}
\begin{proof}
$T_\tau$ is an involution, and it is symplectic because $G\perp G^\perp$ and the form is the orthogonal sum of the forms of the two copies. Locality follows from that of $\pi_G$. Both $\tau_{\cB_G}$ and $\Theta_{T_\tau}$ are automorphisms of $\cA^{\otimes2}$ which send $W(g)^{(1)}\mapsto W(g)^{(2)}$, $W(g)^{(2)}\mapsto W(g)^{(1)}$ and $W(h)^{(s)}\mapsto W(h)^{(s)}$ for $g\in G$, $h\in G^\perp$; since these operators generate $\cA^{\otimes2}$, the two automorphisms coincide.
\end{proof}

\subsection{Separated Clifford QCA and their index}\label{subsec:linear}
Write $\Pau=\Pau_X\oplus\Pau_Z$ for the decomposition of a Pauli vector into its $X$- and $Z$-components identified with the space $\bigoplus_x\F_p^{n_x}$. Let $\langle(a,b),(a',b')\rangle=a\cdot b'-b\cdot a'$, where $a\cdot b=\sum_{x,i}a_{x,i}b_{x,i}$. For a linear map $Y\colon\Pau_X\to\Pau_X$ which is local, let $Y^{*}$ be its transpose with respect to $\cdot$, i.e.\ $Y^{*}b\cdot a=b\cdot Ya$. It is local of the same spread. If $Y$ is a bijection with local inverse, put $Y^{-*}:=(Y^{-1})^{*}=(Y^*)^{-1}$ and
\[
\delta_Y(a,b):=(Ya,\ Y^{-*}b).
\]
Then $\langle\delta_Y(a,b),\delta_Y(a',b')\rangle=Ya\cdot Y^{-*}b'-Y^{-*}b\cdot Ya'=a\cdot b'-b\cdot a'$, so $\delta_Y$ is a symplectic bijection, local with local inverse $\delta_{Y^{-1}}$, and $\Theta_{\delta_Y}$ is a Clifford QCA. Clearly $\delta_{YY'}=\delta_Y\delta_{Y'}$. We also use the on-site symplectic 
\[
\eta(a,b):=(-b,\ a),\qquad\eta^2=-1,
\]
and note that $\eta\,\delta_Y\,\eta^{-1}=\delta_{Y^{-*}}$. The name separated Clifford QCA originates from~\cite{yang2026categorifying}.

Now let the spin system be a chain with $n$ qudits of dimension $p$ in every cell, and let $R=\F_p[x^{\pm1}]$, with $x$ acting on $\Pau_X\cong R^n$ by the translation $i\mapsto i+1$ of the chain (so that $u=\sum_iu_ix^i\in R^n$ is the vector with coordinates $u_i\in\F_p^n$ in the cell $i$). An $R$-linear endomorphism of $R^n$ is a matrix $Y\in\Mat_n(R)$, and its transpose with respect to $\cdot$ is $Y^*:=\bar Y^{\mathsf t}$, where $\bar f(x)=f(x^{-1})$. The units of $R$ are the elements $cx^d$ with $c\in\F_p^\times$, $d\in\Z$, and we write $\deg(cx^d):=d$.

\begin{lemma}\label{lem:delta-index}
For $Y\in GL_n(R)$, the Clifford QCA $\Theta_{\delta_Y}$ of the chain has GNVW index
\[
\ind(\Theta_{\delta_Y})=p^{\deg\det Y}.
\]
\end{lemma}
\begin{proof}
Both sides are homomorphisms $GL_n(R)\to\Qpos$ and $\delta_{YY'}=\delta_Y\delta_{Y'}$). Since $R$ is a Euclidean domain (with Euclidean function $f\mapsto\max\{d:f_d\ne0\}-\min\{d:f_d\ne0\}$ for $f=\sum_df_dx^d$), $SL_n(R)$ is generated by the elementary matrices $1+rE_{ij}$ ($i\ne j$, $r\in R$), so $GL_n(R)$ is generated by these and the diagonal matrices $\operatorname{diag}(c,1,\dots,1)$ and $\operatorname{diag}(x,1,\dots,1)$. It suffices to check the formula on these generators. For $Y=\operatorname{diag}(c,1,\dots,1)$ the QCA $\Theta_{\delta_Y}$ is on-site, so its index is $1=p^0$. For $Y=\operatorname{diag}(x,1,\dots,1)$ we have $Y^{-*}=Y$, so $\Theta_{\delta_Y}$ is the translation $i\mapsto i+1$ of the first qudit of every cell, tensored with the identity on the others, and its index is $p=p^{1}$ by Theorem~\ref{thm:gnvw}(d). For $Y=1+rE_{ij}$ with $r=\sum_dc_dx^d$ we have $Y(a)=a+ra_je_i$ and $Y^{-*}(b)=(1-\bar rE_{ji})b=b-\bar rb_ie_j$. This is the Clifford automorphism given by the product, over all cells $s$ and all $d$, of the generalized controlled-$X$ gates $\mathrm{C}X^{c_d}$ with control the $j$-th qudit of the cell $s$ and target the $i$-th qudit of the cell $s+d$; indeed such a gate maps $X_{\mathrm{ctrl}}\mapsto X_{\mathrm{ctrl}}X_{\mathrm{tgt}}^{c_d}$ and $Z_{\mathrm{tgt}}\mapsto Z_{\mathrm{ctrl}}^{-c_d}Z_{\mathrm{tgt}}$ and fixes $Z_{\mathrm{ctrl}}$ and $X_{\mathrm{tgt}}$. These gates commute with each other, since all controls are $j$-th qudits and all targets are $i$-th qudits with $i\neq j$. Grouping the gates with a fixed $d$ according to $s$ modulo $2|d|+1$ gives finitely many layers of gates with pairwise disjoint supports, and $\Theta_{\delta_Y}$ is their product; so it is a finite-depth circuit and has index $1=p^{\deg\det Y}$ by Theorem~\ref{thm:gnvw}(c).
\end{proof}

\section{The chiral $\Z_p$ subalgebras and the proof of Theorem A}\label{sec:examples}

\subsection{Laurent polynomial notation}\label{subsec:laurent}
Let $p$ be an odd prime and consider $\Z^2$ with $n$ qudits of dimension $p$ per site. Writing $x_1,x_2$ for the translations, the Pauli module $\Pau$ is a free module of rank $2n$ over $R_2=\F_p[x_1^{\pm1},x_2^{\pm1}]$. A vector $u\in R_2^{2n}$ with entries $u_j=\sum_{a\in\Z^2}u_{j,a}x^a$ corresponds to the Pauli vector having coordinates $(u_{1,a},\dots,u_{2n,a})$ at the site $a$, the first $n$ entries being $X$-coordinates and the last $n$ being $Z$-coordinates. For a Laurent polynomial $f$ let $\bar f=f(x_1^{-1},x_2^{-1})$, for a matrix $A$ over $R_2$ let $A^\dagger=\bar A^{\mathsf t}$, and for $u,v\in R_2^{2n}$ put
\[
\lambda(u,v):=u_X^\dagger v_Z-u_Z^\dagger v_X\in R_2,\qquad\text{so that}\qquad\lambda(u,v)=\sum_{a\in\Z^2}\langle u,x^{-a}v\rangle\,x^a .
\]
Thus $\lambda(u,v)=0$ if and only if $u$ is orthogonal to all translates of $v$. A matrix $T\in\Mat_{2n}(R_2)$ defines a translation-invariant linear map of $\Pau$. It is local of spread at most the largest exponent occurring in its entries. For $Y\in\Mat_n(R_2)$ acting on $X$-components, the transpose $Y^*$ of Section~\ref{subsec:linear} is the matrix $Y^\dagger$.

\begin{lemma}\label{lem:matrix-isa}
Let $\cM\in\Mat_n(R_2)$ satisfy $\cM^\dagger=-\cM$ and let $\cM$ be invertible over $R_2$. Let $G\subseteq\Pau$ be the $R_2$-span of the columns of $\binom{\mathbf 1}{\cM}$ and $G'$ the $R_2$-span of the columns of $\binom{\mathbf 1}{-\cM}$. Then $\Pau=G\oplus G'$, $G'=G^\perp$, the projection $\pi_G$ is local of spread at most the largest exponent occurring in $\cM$ and $\cM^{-1}$, and $\cB_G$ is a translation-invariant Pauli invertible subalgebra with commutant $\cB_{G'}$.
\end{lemma}
\begin{proof}
For the $i$-th column $g_i$ of $\binom{\mathbf 1}{\cM}$ and the $j$-th column $g'_j$ of $\binom{\mathbf 1}{-\cM}$ we have $\lambda(g_i,g'_j)=e_i^{\mathsf t}(-\cM)e_j-(\cM e_i)^\dagger e_j=-\cM_{ij}-(\cM^\dagger)_{ij}=0$, so $G\perp G'$. The matrix $H=\begin{pmatrix}\mathbf 1&\mathbf 1\\ \cM&-\cM\end{pmatrix}$ has the inverse $H^{-1}=\frac12\begin{pmatrix}\mathbf 1&\cM^{-1}\\ \mathbf 1&-\cM^{-1}\end{pmatrix}$ over $R_2$, so every $u\in\Pau$ is uniquely $u=\binom{\mathbf 1}{\cM}g+\binom{\mathbf 1}{-\cM}g'$ with $\binom{g}{g'}=H^{-1}u$. Hence $\Pau=G\oplus G'$ and $\pi_G$ is local as stated. If $g\in G\cap G^\perp$, then $g\perp G$ and $g\perp G'$, so $g=0$. If $w\in G^\perp$ with $w=g+g'$, then $g=w-g'\in G\cap G^\perp=0$, so $w\in G'$. Thus $G^\perp=G'$, and Lemma~\ref{lem:pauli-isa} applies.
\end{proof}

In the situation of the lemma, Lemma~\ref{lem:swap-clifford} gives the matrix of the Brauer swap explicitly: writing a vector of $\Pau^{(2)}$ as $(v_{1X},v_{1Z},v_{2X},v_{2Z})$ (the $X$- and $Z$-parts of the two copies) and using $\pi_G\binom{v_X}{v_Z}=\binom{\mathbf 1}{\cM}\cdot\frac12(v_X+\cM^{-1}v_Z)$, where $\binom{\mathbf 1}{\cM}\cdot w:=\binom{w}{\cM w}$ denotes the product of the $2n\times n$ matrix $\binom{\mathbf 1}{\cM}$ with the column vector $w\in R_2^n$,
\begin{equation}\label{eq:tau-closed}
T_\tau=\frac12\begin{pmatrix}
\mathbf 1&-\cM^{-1}&\mathbf 1&\cM^{-1}\\
-\cM&\mathbf 1&\cM&\mathbf 1\\
\mathbf 1&\cM^{-1}&\mathbf 1&-\cM^{-1}\\
\cM&\mathbf 1&-\cM&\mathbf 1
\end{pmatrix}.
\end{equation}
One checks directly from $\cM^\dagger=-\cM$ that $T_\tau^2=1$ and $T_\tau^\dagger\Lambda T_\tau=\Lambda$ for the matrix $\Lambda$ of the form $\lambda$ on $\Pau^{(2)}$.

\subsection{The chiral $\Z_p$ subalgebras}
\begin{definition}\label{def:Zp}
Let $p$ be an odd prime, $k\in\F_p^\times$, and consider $\Z^2$ with two qudits per site, thought of as living on the horizontal edge ($j=1$) and the vertical edge ($j=2$) emanating from the site. Put
\begin{equation}\label{eq:M}
\cM:=\frac k2\,N,\qquad N:=\begin{pmatrix} x_2-x_2^{-1} & x_1^{-1}x_2+x_1^{-1}+x_2-1\\[2pt] -x_1x_2^{-1}-x_1-x_2^{-1}+1 & x_1^{-1}-x_1\end{pmatrix}\in\Mat_2(R_2),
\end{equation}
and let $\cB_{p,k}:=\cB_G\subseteq\cA(\Z^2,p^2)$ with $G$ the $R_2$-span of the columns of $\binom{\mathbf 1}{\cM}$.
\end{definition}

This is the invertible subalgebra generated by the operators $A^{(k)}_e$ of \cite[\S V.B]{SYWTC}. The subalgebra $\cB_{p,k}$ hosts the abelian anyon theory denoted $\Z_p^{(k)}$ in \cite{SYWTC}, with $p$ anyon types. For $p=3$ it hosts a chiral $\Z_3$ theory, like the example of \cite[\S5]{Haah-ISA}, although the two subalgebras are different (Remark~\ref{rem:haah-z3}).

\begin{lemma}\label{lem:Zp-isa}
$\cM^\dagger=-\cM$, and $\cM$ is invertible over $R_2$ with
\begin{equation}\label{eq:Minv}
\cM^{-1}=\frac1{2k}\,N',\qquad N':=\begin{pmatrix} x_1^{-1}-x_1 & -(x_1^{-1}x_2+x_1^{-1}+x_2-1)\\[2pt] x_1x_2^{-1}+x_1+x_2^{-1}-1 & x_2-x_2^{-1}\end{pmatrix},
\end{equation}
where $NN'=N'N=4\cdot\mathbf 1$.
Consequently $\cB_{p,k}$ is a translation-invariant Pauli invertible subalgebra of $\cA(\Z^2,p^2)$ of spread $1$ with commutant $\cB_{p,-k}$, and its Brauer swap is the Clifford QCA with matrix \eqref{eq:tau-closed}, of spread $1$.
\end{lemma}
\begin{proof}
Each diagonal entry of $N$ changes sign under $f\mapsto\bar f$, and $\overline{N_{21}}=-x_1^{-1}x_2-x_1^{-1}-x_2+1=-N_{12}$; hence $N^\dagger=-N$ and $\cM^\dagger=-\cM$. The identity $NN'=4\cdot\mathbf 1$ is a direct multiplication of Laurent polynomials. The remaining assertions are Lemma~\ref{lem:matrix-isa}.
\end{proof}

\begin{remark}[The three-dimensional QCA]\label{rem:3dqca}
By Construction~\ref{constr:pumping}, the pumping QCA $\alpha_{p,k}:=\alpha_{\cB_{p,k}}$ of $\cA(\Z^3,p^2)$ is the Clifford QCA which, on Pauli vectors $g+h$ ($g\in G$, $h\in G^\perp$) of a sheet, acts by $g+h\mapsto zg+h$ with $z$ the translation along the pumping axis.
\end{remark}

\begin{remark}[Comparison with {\cite[\S5]{Haah-ISA}}]\label{rem:haah-z3}
The chiral $\Z_3$ subalgebra $\cA_{\mathrm H}\subseteq\cA(\Z^2,3^2)$ of \cite[\S5]{Haah-ISA} is not $\cB_{3,k}$: it is generated by the translates of two Pauli operators whose Pauli vectors $v_1,v_2$ have commutation matrix $\big(\lambda(v_i,v_j)\big)_{i,j}=\Xi$, the antihermitian matrix of \cite[eq.~(31)]{Haah-ISA}, whereas the generators of $\cB_{3,k}$ have commutation matrix $2\cM=kN$. A direct comparison gives
\[
\Xi(x_1,x_2)=N'(-x_1^{-1},x_2)
\]
(with other conventions for $\lambda$ or for the translations this changes by a reflection and a sign, which does not affect what follows). Nevertheless Theorem~\ref{thm:main} applies to $\cA_{\mathrm H}$. If the Pauli vectors of the generators of two Pauli subalgebras form $R_2$-bases of their spans, with commutation matrices $Q$ and $P^\dagger QP$ for some $P\in GL_n(R_2)$, then $P$ induces a local symplectic bijection between the spans with local inverse, hence, by \eqref{eq:weyl}, a locality-preserving isomorphism between the subalgebras. Now $(2N^{-1})^\dagger(-N)(2N^{-1})=4N^{-1}=N'$; the substitution $x_1\mapsto-x_1$ is induced by the on-site symplectic map that multiplies the Pauli vectors at the sites with odd $x_1$ by $-1$; and $x_1\mapsto x_1^{-1}$ is induced by the reflection $r(a_1,a_2)=(-a_1,a_2)$ of the lattice. Hence $\cA_{\mathrm H}$ is isomorphic, by a locality-preserving isomorphism, to $r(\cB_{3,-1})$, and $[\cA_{\mathrm H}]=[r(\cB_{3,-1})]$ in $\Br_2$. Since $r$ preserves $\{x_2\ge0\}$, conjugating a truncation $t$ of $\tau_{\cB_{3,-1}}$ by $r$ gives a truncation of $\tau_{r(\cB_{3,-1})}$ whose square is the reflection of $t^2$, with GNVW index $\ind(t^2)^{-1}$. Thus $I(r(\cB_{3,-1}))=-I(\cB_{3,-1})=[3]\ne0$ by Theorem~\ref{thm:computation}, so $[\cA_{\mathrm H}]\ne0$ by Proposition~\ref{prop:Brauer-invariance}, and by Proposition~\ref{prop:nontrivial-from-Brauer} the pumping QCA of $\cA_{\mathrm H}$ is not stably a circuit modulo translations.
\end{remark}

\subsection{Reduction to one copy}\label{subsec:one-copy}
For an invertible $\cM'\in\Mat_n(R_2)$ with $\cM'^\dagger=-\cM'$ put
\[
E_{\cM'}:=\begin{pmatrix}0&-\cM'^{-1}\\-\cM'&0\end{pmatrix}\in\Mat_{2n}(R_2).
\]
In the notation of Section~\ref{subsec:linear}, $E_{\cM'}=\delta_{\cM'^{-1}}\circ\eta$, because $(\cM'^{-1})^{-*}=\cM'^\dagger=-\cM'$. Then $E_{\cM'}$ is a symplectic involution of $\Pau$ and $\Theta_{E_{\cM'}}$ is an involutive Clifford QCA. Note that $E_{\cM'}$ is $-1$ on the span of the columns of $\binom{\mathbf 1}{\cM'}$ and $+1$ on the span of the columns of $\binom{\mathbf 1}{-\cM'}$.

\begin{lemma}\label{lem:one-copy}
In the situation of Lemma~\textup{\ref{lem:matrix-isa}}, let $D\colon\Pau\to\Pau$, $D(v_X,v_Z)^{\mathsf t}=(\frac12v_X,v_Z)^{\mathsf t}$, and define the on-site linear map
\[
L\colon\Pau^{(2)}\to\Pau^{(2)},\qquad L(v_1,v_2)=\big(D(v_1+v_2),\,D(v_1-v_2)\big).
\]
Then $L$ is symplectic and $L\,T_\tau\,L^{-1}=\id_\Pau\oplus E_{2\cM}$.
\end{lemma}
\begin{proof}
Since $\langle Dv,Dw\rangle=\frac12\langle v,w\rangle$, we get $\langle L(v_1,v_2),L(w_1,w_2)\rangle=\frac12\big(\langle v_1+v_2,w_1+w_2\rangle+\langle v_1-v_2,w_1-w_2\rangle\big)=\langle v_1,w_1\rangle+\langle v_2,w_2\rangle$; moreover $L$ is invertible since $p$ is odd, with $L^{-1}(u_1,u_2)=\big(\tfrac12D^{-1}(u_1+u_2),\,\tfrac12D^{-1}(u_1-u_2)\big)$. Write $v_s=g_s+h_s$ with $g_s\in G$, $h_s\in G^\perp$. Then $T_\tau(v_1,v_2)=(g_2+h_1,g_1+h_2)$ has the same sum $v_1+v_2$ and the difference $-(g_1-g_2)+(h_1-h_2)=\varepsilon(v_1-v_2)$, where $\varepsilon:=1-2\pi_G$. From $\pi_G\binom{v_X}{v_Z}=\binom{\mathbf 1}{\cM}\frac12(v_X+\cM^{-1}v_Z)$ we get $\varepsilon(v_X,v_Z)^{\mathsf t}=(-\cM^{-1}v_Z,-\cM v_X)^{\mathsf t}$, hence $D\varepsilon D^{-1}(v_X,v_Z)^{\mathsf t}=(-\frac12\cM^{-1}v_Z,-2\cM v_X)^{\mathsf t}=E_{2\cM}(v_X,v_Z)^{\mathsf t}$. Therefore $LT_\tau=(\id_\Pau\oplus E_{2\cM})L$.
\end{proof}

\begin{corollary}\label{cor:one-copy}
In the situation of Lemma~\textup{\ref{lem:one-copy}}, let $\rho$ be a symplectic bijection of $\Pau$, local with local inverse, such that $\Theta_\rho$ is a truncation of $\Theta_{E_{2\cM}}$ to $\{x_2\ge0\}$ and $\Theta_{\rho^2}\in\QCA(L)$. Then $t:=\Theta_L^{-1}(\id\otimes\Theta_\rho)\Theta_L$ is a truncation of $\tau_{\cB_G}$ to $\{x_2\ge0\}$, and $\ind(t^2)=\ind(\Theta_{\rho^2})$.
\end{corollary}
\begin{proof}
Since $\Theta$ is multiplicative, $\Theta_L^{-1}(\id\otimes\Theta_{E_{2\cM}})\Theta_L=\Theta_{L^{-1}(1\oplus E_{2\cM})L}=\Theta_{T_\tau}=\tau_{\cB_G}$ by Lemmas~\ref{lem:swap-clifford} and~\ref{lem:one-copy}. As $\Theta_L$ is on-site it preserves supports, so $t$ agrees with $\tau_{\cB_G}$ and with the identity wherever $\id\otimes\Theta_\rho$ agrees with $\id\otimes\Theta_{E_{2\cM}}$ and with the identity. Finally $t^2=\Theta_L^{-1}(\id\otimes\Theta_{\rho^2})\Theta_L$, and restricted to a strip around $L$ the on-site automorphism $\Theta_L$ is a QCA of the chain of columns, so $\ind(t^2)=\ind(\id\otimes\Theta_{\rho^2})=\ind(\Theta_{\rho^2})$ by Theorem~\ref{thm:gnvw}(b),(e).
\end{proof}

For $\cB_{p,k}$ we have $2\cM=kN$ and $(kN)^{-1}=(4k)^{-1}N'$, so the relevant involution is
\begin{equation}\label{eq:E}
E:=E_{kN}=\begin{pmatrix}0&-(4k)^{-1}N'\\-kN&0\end{pmatrix}.
\end{equation}
We now construct a truncation of $\Theta_E$ by hand.

\subsection{A Toeplitz operator}\label{subsec:toeplitz}
Let $R=\F_p[x_1^{\pm1}]$ and put
\[
\alpha:=1+x_1,\qquad\beta:=1-x_1,\qquad\text{so that}\qquad\alpha+\beta=2,\quad\alpha^2-\beta^2=4x_1,\quad\bar\alpha=x_1^{-1}\alpha .
\]
Let $F:=R[x_2]^2$, the module of vectors $v=\sum_{j\ge0}v_jx_2^j$ with $v_j=(v_{j,1},v_{j,2})^{\mathsf t}\in R^2$. In the Laurent notation of Section~\ref{subsec:laurent}, $F$ is the space of $X$-components (or of $Z$-components) of the Pauli vectors of $\cA(\Z^2,p^2)$ supported in the upper half-plane $\{x_2\ge0\}$; $v_j$ describes the row $x_2=j$, and its two entries the two qudits of a site. For $m\ge0$ let $F_{<m}:=\bigoplus_{j<m}x_2^jR^2$ be the submodule of vectors supported in the rows $0\le x_2<m$, so that $F=F_{<m}\oplus x_2^mF$. For $f,g\in R[x_2^{\pm1}]^2$ with finite support let
\[
\langle f|g\rangle:=\text{the coefficient of $x_2^0$ in $f^\dagger g$}=\sum_j\bar f_j^{\mathsf t}g_j\ \in R .
\]
It is sesquilinear, $\langle rf|g\rangle=\bar r\langle f|g\rangle$ and $\langle f|rg\rangle=r\langle f|g\rangle$ for $r\in R$, satisfies $\langle g|f\rangle=\overline{\langle f|g\rangle}$, and its constant term in $x_1$ is the pairing $f\cdot g$ of Section~\ref{subsec:linear}. Consequently, for an $R$-linear map $Y\colon F\to F$ of finite propagation (that is, local in the sense of Section~\ref{sec:pauli}), the transpose $Y^*$ of Section~\ref{subsec:linear} is the $R$-linear map characterized by $\langle Y^*f|g\rangle=\langle f|Yg\rangle$. The decomposition $F=F_{<m}\oplus x_2^mF$ is orthogonal for $\langle\,\cdot\,|\,\cdot\,\rangle$. Let $e:=(1,0)^{\mathsf t}\in F$ (the first qudit of the site $(0,0)$, as an $X$- or $Z$-component), so that $\langle e|v\rangle=v_{0,1}$, let $\pi\colon R[x_2^{\pm1}]^2\to F$ discard the negative powers of $x_2$, and define
\[
T\colon F\to F,\qquad T(v):=\pi(N'v),
\]
with $N'$ as in \eqref{eq:Minv}.

\begin{lemma}\label{lem:toeplitz}
\begin{enumerate}[label=\textup{(\alph*)}]
\item For $v\in F$ and $j\ge0$, with $v_{-1}:=0$,
\[
T(v)_j=\begin{pmatrix}x_1^{-1}\big(\alpha\beta\,v_{j,1}-\beta\,v_{j,2}-\alpha\,v_{j-1,2}\big)\\ -\beta\,v_{j,1}+\alpha\,v_{j+1,1}-v_{j+1,2}+v_{j-1,2}\end{pmatrix}.
\]
Moreover $T(v)=N'v$ for $v\in x_2F$, and $T^*=-T$.
\item $T$ is injective, and $T(F)=\{w\in F:\ w_{0,1}\in\beta R\}$. More precisely, if $w\in F$ and $w_{0,1}=\beta r$ with $r\in R$, then $T(v)=w$ for $v:=\frac14\big(\pi(Nw)+\alpha r\,e\big)$.
\end{enumerate}
\end{lemma}
\begin{proof}
(a) From \eqref{eq:M} and \eqref{eq:Minv}, $N'=N'_{-1}x_2^{-1}+N'_0+N'_1x_2$ and $N=N_{-1}x_2^{-1}+N_0+N_1x_2$ with
\[
\begin{gathered}
N'_1=\begin{pmatrix}0&-x_1^{-1}\alpha\\0&1\end{pmatrix},\qquad N'_0=\begin{pmatrix}x_1^{-1}\alpha\beta&-x_1^{-1}\beta\\-\beta&0\end{pmatrix},\qquad N'_{-1}=\begin{pmatrix}0&0\\ \alpha&-1\end{pmatrix},\\
N_1=\begin{pmatrix}1&x_1^{-1}\alpha\\0&0\end{pmatrix},\qquad N_0=\begin{pmatrix}0&x_1^{-1}\beta\\ \beta&x_1^{-1}\alpha\beta\end{pmatrix},\qquad N_{-1}=\begin{pmatrix}-1&0\\ -\alpha&0\end{pmatrix},
\end{gathered}
\]
and $(N'v)_j=N'_{-1}v_{j+1}+N'_0v_j+N'_1v_{j-1}$, which is the displayed formula. If $v\in x_2F$, then $N'v$ has no negative powers of $x_2$. For $f,g\in F$, the part of $N'f$ discarded by $\pi$ pairs to zero with $g$, so $\langle Tf|g\rangle=\langle N'f|g\rangle$, which is the coefficient of $x_2^0$ in $f^\dagger N'^\dagger g=-f^\dagger N'g$; thus $\langle Tf|g\rangle=-\langle f|N'g\rangle=-\langle f|Tg\rangle$.

(b) For $v\in F$ put $c:=\alpha v_{0,1}-v_{0,2}$. By (a), the only negative power of $x_2$ in $N'v$ is $x_2^{-1}N'_{-1}v_0$, so
\[
N'v=T(v)+c\,x_2^{-1}\binom01,\qquad\text{and}\qquad T(v)_{0,1}=x_1^{-1}\beta\,c .
\]
In particular $T(F)\subseteq\{w:w_{0,1}\in\beta R\}$. If $T(v)=0$, then $v=\frac14NN'v=\frac c4N\,x_2^{-1}(0,1)^{\mathsf t}$, whose coefficient of $x_2^{-1}$ is $\frac c4N_0(0,1)^{\mathsf t}=\frac c4x_1^{-1}\beta(1,\alpha)^{\mathsf t}$; since $v\in F$ and $R$ is a domain, $c=0$ and hence $v=0$. Conversely, let $w_{0,1}=\beta r$ and put $w':=w+x_1r\,x_2^{-1}(0,1)^{\mathsf t}$. The coefficient of $x_2^{-2}$ in $Nw'$ is $N_{-1}(0,x_1r)^{\mathsf t}=0$, and that of $x_2^{-1}$ is $N_{-1}w_0+N_0(0,x_1r)^{\mathsf t}=(-\beta r+\beta r)(1,\alpha)^{\mathsf t}=0$. Hence $\frac14Nw'\in F$; since the constant coefficient of $x_2^{-1}N(0,x_1r)^{\mathsf t}$ is $N_1(0,x_1r)^{\mathsf t}=\alpha r\,e$, we get $\frac14Nw'=v$. Finally $N'v=w'$, so $T(v)=\pi(w')=w$.
\end{proof}

Thus $T$ is injective with cokernel $F/T(F)\cong R/\beta R\cong\F_p$. If $T$ were bijective, the construction below would work with $T$ itself and would produce an \emph{involutive} truncation (see Remark~\ref{rem:why}). Instead we normalize the defective coordinate $v_{0,1}$. Let $B\colon F\to F$ be the $R$-linear map that multiplies this coordinate by $\beta$, $B(v):=v-x_1v_{0,1}\,e$, and define the $R$-linear map $A\colon F\to F$ by
\begin{equation}\label{eq:A}
A(v):=T(v)+(\alpha v_{0,1}-v_{0,2})\,e .
\end{equation}
Thus $A-T$ is supported in the row $x_2=0$, both in its input and in its output.

\begin{lemma}\label{lem:normalized}
\begin{enumerate}[label=\textup{(\alph*)}]
\item $T=BA$, and $A$ is an $R$-linear bijection of $F$ with $A(v)=N'v$ for $v\in x_2F$ and
\begin{equation}\label{eq:Ainv}
A^{-1}(z)=\frac14\left(\pi(Nz)+z_{0,1}\begin{pmatrix}\alpha-x_1x_2\\-x_1\beta\end{pmatrix}\right)\qquad(z\in F).
\end{equation}
In particular $A^{-1}(z)=\frac14Nz$ for $z\in x_2F$, and $A$ and $A^{-1}$ have finite propagation.
\item $A^*(f)=-T(f)+\langle e|f\rangle\,g$ for $f\in F$, where $g:=(x_1^{-1}\alpha,-1)^{\mathsf t}$.
\item $Ae=(x_1^{-1}\alpha,-\beta)^{\mathsf t}$ and $A^{-*}e=\frac14(Ae-x_2e)$, where $A^{-*}:=(A^{-1})^*=(A^*)^{-1}$. In particular $Ae,A^{-*}e\in F_{<2}$.
\end{enumerate}
\end{lemma}
\begin{proof}
(a) Let $c:=\alpha v_{0,1}-v_{0,2}$. By the proof of Lemma~\ref{lem:toeplitz}(b), $A(v)_{0,1}=(x_1^{-1}\beta+1)c=x_1^{-1}c$, so $B(A(v))=A(v)-c\,e=T(v)$. Since $B$ is injective and $B(F)=\{w:w_{0,1}\in\beta R\}=T(F)$, $A$ is injective, as $T$ is, and surjective: for $z\in F$ write $Bz=T(v)$; then $B(A(v))=Bz$, so $A(v)=z$. If $v\in x_2F$, then $c=0$ and $A(v)=T(v)=N'v$. Given $z\in F$, the vector $w:=Bz$ has $w_{0,1}=\beta z_{0,1}$, so by Lemma~\ref{lem:toeplitz}(b) the vector $v:=\frac14(\pi(Nw)+\alpha z_{0,1}e)$ satisfies $B(A(v))=T(v)=Bz$, i.e.\ $A(v)=z$. Since $\pi(Ne)=(x_2,\beta)^{\mathsf t}$, we have $\pi(Nw)=\pi(Nz)-x_1z_{0,1}(x_2,\beta)^{\mathsf t}$, which gives \eqref{eq:Ainv}. If $z\in x_2F$, then $z_{0,1}=0$ and $Nz\in F$. Finally, $A$ and $A^{-1}$ map $x_2^jR^2$ into $x_2^{j-1}R^2\oplus x_2^jR^2\oplus x_2^{j+1}R^2$ (where $x_2^{-1}R^2:=0$) by matrices with entries in a fixed finite-dimensional space of Laurent polynomials, so they have finite propagation.

(b) Since $\overline{x_1^{-1}\alpha}=x_1\bar\alpha=\alpha$, we have $\langle g|v\rangle=\alpha v_{0,1}-v_{0,2}$, so $A=T+H$ with $H(v):=\langle g|v\rangle\,e$. From $\langle f|Hv\rangle=\langle g|v\rangle\langle f|e\rangle=\big\langle\langle e|f\rangle g\big|v\big\rangle$ we get $H^*(f)=\langle e|f\rangle\,g$, and $T^*=-T$ by Lemma~\ref{lem:toeplitz}(a).

(c) By Lemma~\ref{lem:toeplitz}(a), $T(e)=(x_1^{-1}\alpha\beta,-\beta)^{\mathsf t}$, so $Ae=(x_1^{-1}\alpha\beta+\alpha,-\beta)^{\mathsf t}=(x_1^{-1}\alpha,-\beta)^{\mathsf t}$. The row $0$ of $\pi(Nz)$ is $N_{-1}z_1+N_0z_0$, so \eqref{eq:Ainv} gives
\[
\langle e|A^{-1}z\rangle=\tfrac14\big(\alpha z_{0,1}+x_1^{-1}\beta z_{0,2}-z_{1,1}\big)=\big\langle\tfrac14(Ae-x_2e)\big|z\big\rangle ,
\]
using $\overline{-\beta}=x_1^{-1}\beta$. As $\langle e|A^{-1}z\rangle=\langle A^{-*}e|z\rangle$ for all $z\in F$, this proves the claim.
\end{proof}

\subsection{The truncation}\label{subsec:truncation}
Identify the Pauli vectors of $\cA(\Z^2,p^2)$ supported in $\{x_2\ge0\}$ with pairs $(a,b)\in F\oplus F$ of $X$- and $Z$-components.

\begin{proposition}\label{prop:rho}
Let $k\in\F_p^\times$ and let $E$ be as in \eqref{eq:E}. Define $\rho\colon\Pau\to\Pau$ to be the identity on Pauli vectors supported in $\{x_2<0\}$ and
\[
\rho(a,b):=\big(-(4k)^{-1}\,Ab,\ \ 4k\,A^{-*}a\big)\qquad\text{for }(a,b)\in F\oplus F .
\]
\begin{enumerate}[label=\textup{(\alph*)}]
\item $\rho$ is a symplectic bijection, local with local inverse, and $\Theta_\rho$ is a truncation of $\Theta_E$ to $\{x_2\ge0\}$: it agrees with $\Theta_E$ on operators supported in $\{x_2\ge2\}$ and with the identity on operators supported in $\{x_2<0\}$.
\item $\rho^2$ is the identity on Pauli vectors supported in $\{x_2<0\}$, and $\rho^2(a,b)=(Ca,C^{-*}b)$ for $(a,b)\in F\oplus F$, where $C:=-AA^{-*}$ is an $R$-linear automorphism of $F$ satisfying
\begin{equation}\label{eq:CB}
C(Bz)=z-x_1^{-1}\langle A^{-*}e|z\rangle\,Ae\qquad(z\in F).
\end{equation}
In particular $C$ is the identity on $x_2^2F$ and maps $F_{<2}=R^2\oplus x_2R^2$ into itself.
\end{enumerate}
\end{proposition}
\begin{proof}
(a) Let $Y$ be the linear map of $\Pau_X$ which is $(4k)^{-1}A$ on $F$ and the identity on the $X$-components supported in $\{x_2<0\}$, and let $\eta_+$ be the on-site map which is $\eta$ on the sites of $\{x_2\ge0\}$ and the identity elsewhere. By Lemma~\ref{lem:normalized}(a), $Y$ is a bijection, local with local inverse, and $Y^{-*}=4kA^{-*}$ on $F$. Since $\delta_Y\eta_+(a,b)=\delta_Y(-b,a)=(-Yb,Y^{-*}a)$, we have $\rho=\delta_Y\circ\eta_+$, which is a symplectic bijection, local with local inverse, by Section~\ref{subsec:linear}. Let $a,b\in x_2^2F$. Then $Ab=N'b$ by Lemma~\ref{lem:normalized}(a). Moreover $f:=-\frac14Na$ lies in $x_2F$, so $\langle e|f\rangle=0$ and $T(f)=N'f$, and Lemma~\ref{lem:normalized}(b) gives $A^*f=-N'f=\frac14N'Na=a$, i.e.\ $A^{-*}a=-\frac14Na$. Hence $\rho(a,b)=(-(4k)^{-1}N'b,-kNa)=E(a,b)$. Thus $\Theta_\rho$ and $\Theta_E$ agree on all Weyl operators $W(v)$ with $v$ supported in $\{x_2\ge2\}$, and $\Theta_\rho$ fixes those with $v$ supported in $\{x_2<0\}$; since these Weyl operators span the respective local algebras, $\Theta_\rho$ is a truncation of $\Theta_E$ (Definition~\ref{def:truncation} with $m=0$, $r=2$).

(b) We have $\rho^2(a,b)=\rho(-Yb,Y^{-*}a)=(-YY^{-*}a,-Y^{-*}Yb)$, where $-YY^{-*}=-AA^{-*}=C$ and $-Y^{-*}Y=C^{-*}$ on $F$. Since $T=BA$ and $T^*=-T$ (Lemmas~\ref{lem:toeplitz}(a) and~\ref{lem:normalized}(a)), $A^*B^*=-BA$; hence $B^*A^{-1}=-A^{-*}B$ and
\[
CB=-AA^{-*}B=AB^*A^{-1}.
\]
From $\langle f|Bv\rangle=\langle f|v\rangle-x_1\langle f|e\rangle\langle e|v\rangle$ we get $B^*(v)=v-x_1^{-1}v_{0,1}\,e$, so $AB^*A^{-1}z=z-x_1^{-1}\langle e|A^{-1}z\rangle Ae$, which is \eqref{eq:CB}. By Lemma~\ref{lem:normalized}(c), $Ae$ and $A^{-*}e$ lie in $F_{<2}$. If $z\in x_2^2F$, then $Bz=z$ and $\langle A^{-*}e|z\rangle=0$, so $Cz=z$. If $z\in F_{<2}$, then $C(Bz)\in F_{<2}$ by \eqref{eq:CB}; applied to the $z\in F_{<2}$ with $z_{0,1}=0$ (for which $Bz=z$) and to $z=e$ (for which $Bz=\beta e$), this gives $C(F_{<2})\subseteq F_{<2}$, because $\beta\,C(e)\in F_{<2}$ implies $C(e)\in F_{<2}$.
\end{proof}

\subsection{The index}\label{subsec:index}
\begin{theorem}\label{thm:computation}
Let $p$ be an odd prime and $k\in\F_p^\times$. The Brauer swap of $\cB_{p,k}$ has a truncation $t$ to $\{x_2\ge0\}$ with
\[
\ind(t^2)=p^{-1}.
\]
Consequently $\tau_{\cB_{p,k}}$ is truncatable and $I(\cB_{p,k})=[p]\ne0$.
\end{theorem}
\begin{proof}
Let $\rho$ be as in Proposition~\ref{prop:rho}. By part (b) of that proposition, $C=C_{<2}\oplus\id$ with respect to the orthogonal decomposition $F=F_{<2}\oplus x_2^2F$, where $C_{<2}$ is the restriction of $C$ to $F_{<2}\cong R^4$ (with the basis $e$, $(0,1)^{\mathsf t}$, $x_2e$, $x_2(0,1)^{\mathsf t}$); as $C$ is bijective, $C_{<2}\in GL_4(R)$, and $C^{-*}=C_{<2}^{-*}\oplus\id$. Hence $\rho^2$ is the identity on all Pauli vectors supported outside the rows $0\le x_2\le1$ and preserves the Pauli vectors supported in these rows, so $\Theta_{\rho^2}\in\QCA(L)$, and as a QCA of the chain whose cells are the columns of the rows $0\le x_2\le1$ (four qudits per cell) it is $\Theta_{\delta_{C_{<2}}}$; by Theorem~\ref{thm:gnvw}(e) its index is the same whether computed on these rows or on any larger strip. By Corollary~\ref{cor:one-copy} (with $2\cM=kN$) and Proposition~\ref{prop:rho}(a), $t:=\Theta_L^{-1}(\id\otimes\Theta_\rho)\Theta_L$ is a truncation of $\tau_{\cB_{p,k}}$, and by Lemma~\ref{lem:delta-index}
\[
\ind(t^2)=\ind(\Theta_{\rho^2})=p^{\deg\det C_{<2}} .
\]
In the basis above, $B$ restricts to $B_{<2}=\operatorname{diag}(\beta,1,1,1)$, and by \eqref{eq:CB}, $C_{<2}B_{<2}=\mathbf 1_4+uw^{\mathsf t}$, where $u\in R^4$ is the coordinate vector of $-x_1^{-1}Ae$ and $w\in R^4$ is obtained from the coordinate vector of $A^{-*}e$ by applying $f\mapsto\bar f$ to each entry, so that $w^{\mathsf t}z=\langle A^{-*}e|z\rangle$. By the identity $\det(\mathbf 1+uw^{\mathsf t})=1+w^{\mathsf t}u$, valid over any commutative ring,
\[
\beta\,\det C_{<2}=1-x_1^{-1}\langle A^{-*}e|Ae\rangle=1-x_1^{-1}\langle e|A^{-1}Ae\rangle=1-x_1^{-1},
\]
so $\det C_{<2}=(1-x_1^{-1})/(1-x_1)=-x_1^{-1}$. Thus $\deg\det C_{<2}=-1$ and $\ind(t^2)=p^{-1}$. Since $\cB_{p,k}$ is translation invariant, $\tau_{\cB_{p,k}}$ is truncatable (Remark~\ref{rem:ti-truncatable}), and $I(\cB_{p,k})=[p^{-1}]=[p]$ by Proposition~\ref{prop:well-defined}; this class is nonzero because a prime is not the square of a rational number.
\end{proof}
\begin{remark}[Where the anyons enter]\label{rem:why}
If $T$ were bijective, we could take $B=1$ and $A=T$; then $C=-TT^{-*}=1$, so $\rho$ would be an involution and $\ind(t^2)=1$. The index thus comes from the cokernel $F/T(F)\cong R/\beta R\cong\F_p$ of $T$, whose dimension $1$ is $\log_p$ of the number of anyons of $\Z_p^{(k)}$ (see Section~\ref{subsec:parity} and Remark~\ref{rem:anyons}).
\end{remark}
\subsection{A cokernel-parity formula}\label{subsec:parity}
The computation above is an instance of a general formula, which we record although it is not needed for Theorem~A. In this subsection $n\ge1$ is arbitrary, $F:=R[x_2]^n$, and $\pi$, $\langle\,\cdot\,|\,\cdot\,\rangle$, the transpose $Y^*$ of an $R$-linear map $Y\colon F\to F$ of finite propagation, and $F_{<m}:=\bigoplus_{j<m}x_2^jR^n$ are as in Section~\ref{subsec:toeplitz}. An $R$-linear map $H\colon F\to F$ is \emph{supported in $F_{<m}$} if $H(x_2^mF)=0$ and $H(F)\subseteq F_{<m}$; then so is $H^*$. If $X-1$ is supported in $F_{<m}$, then $X=X_{<m}\oplus\id$ with respect to $F=F_{<m}\oplus x_2^mF$, with $X_{<m}\in\Mat_{nm}(R)$, and we put $\det_\partial X:=\det X_{<m}$, which does not depend on $m$; if $X$ is bijective, then $X_{<m}\in GL_{nm}(R)$ and $\det_\partial X\in R^\times$. More generally, for an $\F_p(x_1)$-linear endomorphism $X$ of $\F_p(x_1)\otimes_RF$ such that $X-1$ has finite rank, let $\det X$ be the determinant of the restriction of $X$ to any finite-dimensional subspace $W$ containing the image of $X-1$; such a $W$ is $X$-invariant, and enlarging $W$ adds an identity block on the quotient, so $\det X$ is well defined. This determinant is multiplicative (for $XX'$ take $W$ containing the images of $X-1$ and $X'-1$), invariant under conjugation by $\F_p(x_1)$-linear automorphisms $g$ (replace $W$ by $gW$), and equal to $\det_\partial X$ if $X-1$ is supported in $F_{<m}$ (take $W=\F_p(x_1)\otimes_RF_{<m}$). For $0\ne f=\sum_{i=l}^uf_ix_1^i\in R$ with $f_lf_u\ne0$ put $\width(f):=u-l$.

\begin{theorem}\label{thm:parity}
Let $T\colon F\to F$ be an $R$-linear map of finite propagation with $T^*=-T$ and $\ker T=0$. Let $U\colon F\to F$ be an $R$-linear bijection such that $U$ and $U^{-1}$ have finite propagation and $U-T$ is supported in $F_{<m}$ for some $m$. Then $TU^{-1}-1$ and $-UU^{-*}-1$ are supported in $F_{<m'}$ for some $m'\ge m$, $a:=\det_\partial(TU^{-1})$ is nonzero, and
\[
\det\nolimits_\partial(-UU^{-*})=\bar a/a .
\]
Moreover $\dim_{\F_p}\coker T=\width(a)<\infty$, and $\deg\det_\partial(-UU^{-*})\equiv\dim_{\F_p}\coker T\pmod2$.
\end{theorem}
\begin{proof}
Put $B:=TU^{-1}$ and $C:=-UU^{-*}$, and choose $s\ge0$ such that $U^{\pm1}$, hence also $U^*$ and $U^{-*}$, map each $x_2^jR^n$ into $\bigoplus_{|i-j|\le s}x_2^iR^n$. Since $B-1=-(U-T)U^{-1}$, the map $B-1$ is supported in $F_{<m+s}$, and $B$ is injective because $T$ is. Hence $a\ne0$, and $\coker T=F/B(F)\cong\coker B_{<m+s}$. By the Smith normal form over the principal ideal domain $R$, $\coker B_{<m+s}\cong\bigoplus_iR/(d_i)$ with $\prod_id_i=a$ up to a unit. Each $d_i$ is a unit times a polynomial $d_i'\in\F_p[x_1]$ with $d_i'(0)\ne0$ and $\deg d_i'=\width(d_i)$, and $x_1$ is invertible modulo $d_i'$, so $R/(d_i)\cong\F_p[x_1]/(d_i')$ has dimension $\width(d_i)$. As $\width$ is additive under products, $\dim_{\F_p}\coker T=\width(a)$.

From $T=BU$ and $T^*=-T$ we get $U^*B^*=-BU$, hence $-U^{-*}B=B^*U^{-1}$ and
\[
CB=UB^*U^{-1}.
\]
Since $B^*-1$ is supported in $F_{<m+s}$, the map $UB^*U^{-1}-1=U(B^*-1)U^{-1}$ is supported in $F_{<m'}$ with $m':=m+2s$. Both $B$ and $UB^*U^{-1}$ are the identity on $x_2^{m'}F$, hence so is $C$. For $z\in F_{<m'}$ we have $C(Bz)=UB^*U^{-1}z\in F_{<m'}$; as $aF_{<m'}\subseteq B(F_{<m'})$ by Cramer's rule, $aC(z)\in F_{<m'}$, hence $C(z)\in F_{<m'}$. Thus $C-1$ is supported in $F_{<m'}$, and $\det_\partial C\in R^\times$. Extending scalars to $\F_p(x_1)$ and taking determinants in $CB=UB^*U^{-1}$, which is conjugate to $B^*$ by $U$, gives $\det_\partial C\cdot a=\det_\partial B^*=\bar a$. Finally write $\det_\partial C=cx_1^j$ with $c\in\F_p^\times$ and $a=\sum_{i=l}^ua_ix_1^i$ with $a_la_u\ne0$; comparing the lowest exponents in $\bar a=cx_1^ja$ gives $-u=j+l$, so $j=-l-u\equiv u-l=\width(a)\pmod2$.
\end{proof}

\begin{proposition}\label{prop:repair}
Let $S\in\Mat_n(R_2)$ be invertible over $R_2$ with $S^\dagger=-S$, and let $T:=\pi S|_F$, an $R$-linear map of finite propagation with $T^*=-T$ \textup(as in the proof of Lemma~\textup{\ref{lem:toeplitz}(a))}. Suppose that $T$ is injective and $\dim_{\F_p}\coker T<\infty$, and let $m\ge0$ be such that $S^{-1}(x_2^mF)\subseteq F$. Then $L:=T(F)\cap F_{<m}$ is a free $R$-module of rank $nm$. If the columns of $B_0\in\Mat_{nm}(R)$ form an $R$-basis of $L$ and $B:=B_0\oplus\id$, then $B$ is injective with $B(F)=T(F)$, the map $A\colon F\to F$ defined by $BA=T$ satisfies the hypotheses on $U$ in Theorem~\textup{\ref{thm:parity}}, and $TA^{-1}=B$. In particular $\det_\partial(-AA^{-*})=\bar\delta/\delta$ with $\delta:=\det B_0$, and $\dim_{\F_p}\coker T=\width(\delta)$.
\end{proposition}
\begin{proof}
For $z\in x_2^mF$ we have $S^{-1}z\in F$ and $T(S^{-1}z)=z$; hence $x_2^mF\subseteq T(F)$ and $T(F)=L\oplus x_2^mF$. Since $F_{<m}/L\cong F/T(F)$ is finite-dimensional over $\F_p$, it is a torsion $R$-module, so the submodule $L$ of $F_{<m}\cong R^{nm}$ is free of rank $nm$. Thus $B$ is injective with $B(F)=T(F)$, and as in the proof of Lemma~\ref{lem:normalized}(a), $A$ is a well-defined $R$-linear bijection. Choose $s\ge0$ such that $T$ and $S^{-1}$ map each $x_2^jR^n$ into $\bigoplus_{|i-j|\le s}x_2^iR^n$. If $T(v)=l+w$ with $l\in L$ and $w\in x_2^mF$, then $A(v)=B_0^{-1}l+w$, so $A(v)-T(v)=B_0^{-1}l-l\in F_{<m}$, and $l=0$ if $v\in x_2^{m+s}F$; hence $A-T$ is supported in $F_{<m+s}$, and $A$ has finite propagation. If $z\in x_2^{m+2s}F$, then $v:=S^{-1}z\in x_2^{m+s}F$ and $A(v)=T(v)=z$, so $A^{-1}z=S^{-1}z$; since $A^{-1}$ is determined on $F_{<m+2s}$ by its values on finitely many basis vectors, $A^{-1}$ has finite propagation. Finally $TA^{-1}=B$, and Theorem~\ref{thm:parity} applies with $a=\det_\partial B=\delta$.
\end{proof}

\begin{corollary}\label{cor:parity}
Let $\cM$ be as in Lemma~\textup{\ref{lem:matrix-isa}}, and suppose that $T:=\pi(2\cM)^{-1}|_F$ is injective with $\dim_{\F_p}\coker T<\infty$. Then $\tau_{\cB_G}$ is truncatable and
\[
I(\cB_G)=\big[p^{\dim_{\F_p}\coker T}\big].
\]
The same holds with $T$ replaced by $T':=\pi(2\cM)|_F$. In particular, if both $T$ and $T'$ are injective with finite-dimensional cokernels, then $\dim_{\F_p}\coker T\equiv\dim_{\F_p}\coker T'\pmod2$.
\end{corollary}
\begin{proof}
Let $U:=A$ be as in Proposition~\ref{prop:repair} for $S:=(2\cM)^{-1}$, choose $m'$ such that $U-T$ and $C-1$ are supported in $F_{<m'}$, where $C:=-UU^{-*}$ (Theorem~\ref{thm:parity}), and choose $s$ such that $S$ and $2\cM$ map each $x_2^jR^n$ into $\bigoplus_{|i-j|\le s}x_2^iR^n$. Let $\rho$ be the identity on Pauli vectors supported in $\{x_2<0\}$ and $\rho(a,b):=(-Ub,U^{-*}a)$ for $(a,b)\in F\oplus F$. As in the proof of Proposition~\ref{prop:rho}, $\rho=\delta_Y\circ\eta_+$ with $Y=U$ on $F$, so $\rho$ is a symplectic bijection, local with local inverse, and $\rho^2(a,b)=(Ca,C^{-*}b)$ on $F\oplus F$. Let $a,b\in x_2^{m'+s}F$. Then $Ub=Tb=Sb$, and $f:=-2\cM a\in x_2^{m'}F$ satisfies $U^*f=-Tf+(U-T)^*f=-Sf=a$, since $(U-T)^*$ is supported in $F_{<m'}$. Hence $\rho(a,b)=(-Sb,-2\cM a)=E_{2\cM}(a,b)$, so $\Theta_\rho$ is a truncation of $\Theta_{E_{2\cM}}$ to $\{x_2\ge0\}$. As $C=C_{<m'}\oplus\id$, the argument in the proof of Theorem~\ref{thm:computation} shows $\Theta_{\rho^2}\in\QCA(L)$ and $\ind(\Theta_{\rho^2})=p^{\deg\det_\partial C}$. By Corollary~\ref{cor:one-copy}, $\tau_{\cB_G}$ has a truncation $t$ with $\ind(t^2)=p^{\deg\det_\partial C}$, and $\deg\det_\partial C\equiv\dim_{\F_p}\coker T\pmod2$ by Theorem~\ref{thm:parity}. Since $\cB_G$ is translation invariant, $\tau_{\cB_G}$ is truncatable (Remark~\ref{rem:ti-truncatable}), and the formula follows from Proposition~\ref{prop:well-defined}.

For $T'$, let $V$ be as in Proposition~\ref{prop:repair} for $S:=2\cM$, and put $\rho'(a,b):=(V^{-*}b,-Va)$ on $F\oplus F$; thus $\rho'=\delta_{Y'}\circ\eta_+$ with $Y'=-V^{-*}$ on $F$. The same argument, with $2\cM$ and $(2\cM)^{-1}$ exchanged, shows that $\Theta_{\rho'}$ is a truncation of $\Theta_{E_{2\cM}}$, and now $\rho'^2(a,b)=(C'a,C'^{-*}b)$ with $C'^{-*}=-VV^{-*}$. Hence $\det_\partial C'=\overline{\det_\partial(-VV^{-*})}^{\,-1}$ has the same degree as $\det_\partial(-VV^{-*})$, which is congruent to $\dim_{\F_p}\coker T'$ modulo $2$ by Theorem~\ref{thm:parity}. The last assertion holds because $I(\cB_G)$ is well defined.
\end{proof}

\subsection{Proof of Theorem A}
\begin{theorem}\label{thm:main}
Let $p$ be an odd prime, $k\in\F_p^\times$ and $\cB=\cB_{p,k}$.
\begin{enumerate}[label=\textup{(\roman*)}]
\item $I(\cB)=[p]\ne0$ in $\Qpos/\Qpos^2$, and $[\cB]\ne0$ in $\Br_2$.
\item The pumping QCA $\alpha_{p,k}$ of $\cA(\Z^3,p^2)$ is not stably a circuit modulo translations. The same holds for every QCA $\alpha$ of a spin system on $\Z^3$ with $\partial\alpha=\pm[\cB]$, in particular for the QCA $\alpha^{(k)}_{\mathrm{ISA}}$ of \cite[eq.~(214)]{SYWTC}.
\item The $\Z/2$-action $\tau_\cB$ on two copies of $\cA(\Z^2,p^2)$ is neither stably nor weakly linearizable.
\end{enumerate}
\end{theorem}
\begin{proof}
(i) is Theorem~\ref{thm:computation}; if $[\cB]$ were zero, $I(\cB)$ would vanish by Proposition~\ref{prop:Brauer-invariance}. (ii) is Proposition~\ref{prop:nontrivial-from-Brauer}. (iii) is Corollary~\ref{cor:onsite}.
\end{proof}

\begin{corollary}\label{cor:inf-gen}
Write $g_{p,k}:=[\cB_{p,k}]\in\Br_2$.
\begin{enumerate}[label=\textup{(\roman*)}]
\item Let $n_{p,k}\in\Z$, almost all zero, and suppose that $\sum_kn_{p,k}$ is odd for some odd prime $p$. Then
\[
\textstyle\sum_{p,k}n_{p,k}\,g_{p,k}\ne0 .
\]
\item If $p\equiv1\pmod4$, then $g_{p,k}$ has order exactly $2$. Hence, for any choice of $k_p\in\F_p^\times$, the classes $g_{p,k_p}$ with $p\equiv1\pmod4$ span a subgroup of $\Br_2$ isomorphic to $\bigoplus_{p\equiv1\,(4)}\Z/2$.
\item $\Br_2$ is not finitely generated, and neither is the QCA group on $\Z^3$.
\end{enumerate}
\end{corollary}
\begin{proof}
The classes represented by invertible subalgebras with truncatable Brauer swap form a subgroup of $\Br_2$ (Lemma~\ref{lem:I-stabilization}(iii) and Proposition~\ref{prop:additive}), on which $I$ is a homomorphism (Theorem~\ref{thm:Brauer-invariance}) to $\Qpos/\Qpos^2$, the $\F_2$-vector space with basis the classes of the primes. Since $I(g_{p,k})=[p]$ (Theorem~\ref{thm:computation}), we get $I\big(\sum n_{p,k}g_{p,k}\big)=\big[\prod_pp^{\sum_kn_{p,k}}\big]$, which proves (i). For (ii), choose $a\in\F_p$ with $a^2=-1$. The on-site symplectic map $(v_X,v_Z)\mapsto(av_X,a^{-1}v_Z)$ maps the graph of $\cM$ onto the graph of $a^{-2}\cM=-\cM$, so the induced on-site automorphism maps $\cB_{p,k}$ onto its commutant $\cB_{p,-k}$; thus $g_{p,k}=-g_{p,k}$, while $g_{p,k}\ne0$ by (i). By (i), no nonempty sum of distinct classes $g_{p,k_p}$ vanishes. (iii) For $\Br_2$ this follows from (ii), since the torsion subgroup of a finitely generated abelian group is finite. For QCA, the argument in the proof of Lemma~\ref{lem:boundary-trivial}(i) gives $\cB(\alpha\otimes\beta,n)=\cB(\alpha,n)\otimes\cB(\beta,n)$, so $\partial$ is additive under tensor products; by Lemma~\ref{lem:boundary-trivial} it vanishes on stabilized circuits composed with translations, so it induces a homomorphism from this group to $\Br_2$, whose image contains $\partial\alpha_{p,k}=g_{p,k}$ (Proposition~\ref{prop:pumping-boundary}) and hence the subgroup of (ii). As that subgroup is not finitely generated, neither is the group of QCA.
\end{proof}

\section{Physical discussion and conjectures}\label{sec:discussion}

Nothing in this section is used elsewhere in the paper. We discuss what the invariant $I$ sees of the anyon theory hosted by an invertible subalgebra, and state the conjectures suggested by this picture.

\subsection{Anyons and the invariant}\label{subsec:anyons}

\begin{remark}[Counting anyons]\label{rem:anyons}
The value $I(\cB_{p,k})=[p]$ is, up to squares, the number of anyons of the theory $\Z_p^{(k)}$ hosted by $\cB_{p,k}$. Equivalently, it is the class of $D^2$, where $D=\sqrt p$ is the total quantum dimension of $\Z_p^{(k)}$. By additivity (Proposition~\ref{prop:additive}) the same holds for tensor products of the $\cB_{p,k}$. In the proof of Theorem~\ref{thm:computation} the exponent of $p$ arises from the one-dimensional cokernel of the half-plane compression $T$ of Section~\ref{subsec:toeplitz} (Remark~\ref{rem:why}), and Corollary~\ref{cor:parity} shows that, for every translation-invariant Pauli invertible subalgebra as in Lemma~\ref{lem:matrix-isa} whose half-plane compression is injective with finite-dimensional cokernel, $I$ is determined by the parity of $\dim_{\F_p}\coker T$. That only a square class appears is to be expected: condensing the bosons of an isotropic subgroup $H$ of an abelian anyon theory $\cM$ produces the Witt-equivalent theory $H^\perp/H$, which has $|\cM|/|H|^2$ anyons.
\end{remark}

\begin{remark}[Classes modulo $2$]\label{rem:mod2}
Since $I$ is additive, $I(\cB\otimes\cB)=2I(\cB)=0$: the invariant sees only the class of $[\cB]$ modulo $2$. In the Clifford Brauer group the class of $\cB_{p,k}$ has order $2$ for $p\equiv1\pmod4$ and order $4$ for $p\equiv3\pmod4$ \cite{Haah-Clifford,SYWTC}. Theorem~\ref{thm:main} shows that it does not become divisible by $2$ in the full Brauer group, in the sense that it is not the class of $\cB'\otimes\cB'$ for any invertible subalgebra $\cB'$ with truncatable Brauer swap.
\end{remark}

\begin{remark}[Vanishing chiral central charge]\label{rem:central-charge}
In the convention of \cite{SYWTC}, $\Z_p^{(k)}$ has topological spins $\theta_a=e^{2\pi ika^2/p}$. Its normalized Gauss sum is $p^{-1/2}\sum_{a\in\Z/p}e^{2\pi ika^2/p}=\big(\tfrac kp\big)\varepsilon_p$, with $\varepsilon_p=1$ for $p\equiv1\pmod4$ and $\varepsilon_p=i$ for $p\equiv3\pmod4$, and by the Gauss--Milgram formula it equals $e^{2\pi ic_-/8}$. Hence for $p\equiv1\pmod4$ and $k$ a square, $c_-\equiv0\pmod8$, although $I(\cB_{p,k})=[p]\ne0$. For instance $\Z_5^{(1)}$ is the discriminant form of the even lattice with Gram matrix $\left(\begin{smallmatrix}2&1\\1&-2\end{smallmatrix}\right)$, of signature $0$. Thus the invariant detects nontrivial QCA classes whose boundary anyon theory has vanishing chiral central charge modulo $8$, and by Corollary~\ref{cor:inf-gen} infinitely many independent such classes.
\end{remark}

\subsection{Conjectures and remarks}\label{subsec:conjectures}

\begin{conjecture}\label{conj:anyons}
Let $\cB$ be an invertible subalgebra with truncatable Brauer swap that hosts an abelian anyon theory $\cM$. Then $I(\cB)=[\,|\cM|\,]$. In other words, $\ind(t^2)$ is, up to squares, the square $D_\cM^2=|\cM|$ of the total quantum dimension of $\cM$. 
\end{conjecture}

In particular, if Conjecture~7.4 applies, it would rule out strictly two-dimensional local
commuting-projector Hamiltonians realizing $\mathbb{Z}_p^{(k)}$ on
tensor-product spin systems with finite-dimensional on-site Hilbert spaces,
since the full local operator algebra $A$ has an on-site Brauer swap and
hence $I(A)=0$, whereas $[|\mathbb{Z}_p^{(k)}|]=[p]\neq 0$ for every odd
prime $p$.

\begin{remark}[Clifford classes]\label{rem:clifford-rank}
Fix an odd prime $p$, and let $\mathfrak C_p$ be the group of translation-invariant Clifford QCA on $\Z^3$ with qudits of dimension $p$, modulo Clifford circuits and shifts. By \cite{Haah-Clifford-cat} there is an isomorphism $\Phi\colon W(\F_p)\to\mathfrak C_p$, where $W(\F_p)$ is the Witt group of nondegenerate quadratic forms over $\F_p$. For such a QCA $\alpha$, let $q_\alpha$ be a quadratic form with $\Phi([q_\alpha])=[\alpha]$. It is determined only up to Witt equivalence, which changes the dimension by an even number, so the parity of $\dim q_\alpha$ is well defined.

The constructions of \cite{SYWTC} suggest that $q_{\alpha_{p,k}}$ can be taken to be one-dimensional, namely the quadratic form of the anyon theory $\Z_p^{(k)}$. Now write $[\alpha]=\sum_kn_k[\alpha_{p,k}]$ in $\mathfrak C_p$, so that $\dim q_\alpha\equiv\sum_kn_k\pmod2$. Hence $\partial\alpha=\sum_kn_k[\cB_{p,k}]$, and Theorem~\ref{thm:computation} together with the additivity of $I$ (Theorem~\ref{thm:Brauer-invariance}) gives
$
I(\partial\alpha)=\big[p^{\dim q_\alpha}\big].
$
Thus, on Clifford QCA, $I$ records exactly the parity of $\dim q_\alpha$. If $\dim q_\alpha$ is odd, then $\alpha$ is not stably a circuit modulo generalized shifts, even when non-Clifford circuits are allowed (Proposition~\ref{prop:nontrivial-from-Brauer} and Remark~\ref{rem:gen-shift}). Since $q_\alpha$ defines an abelian anyon theory with $p^{\dim q_\alpha}$ anyons, this agrees with Remark~\ref{rem:anyons}. Classes with $\dim q_\alpha$ even, such as $2[\alpha_{p,k}]$ for $p\equiv3\pmod4$, are not detected by $I$.
\end{remark}

\begin{remark}[The three-fermion QCA]\label{rem:3F}
We have not included $I(\cB_{3F})$ for the three-fermion subalgebra. A computer search found a truncation of the Brauer swap with $\ind(t^2)=2^{-6}$, which would give $I(\cB_{3F})=0$. The three-fermion theory has $4$ anyons, so Conjecture~\ref{conj:anyons} also predicts $I(\cB_{3F})=[4]=0$. Thus we do not expect our method to say anything about the three-fermion QCA. By contrast, the chiral semion theory~\cite{shirley2022three} has $2$ anyons, so Conjecture~\ref{conj:anyons} predicts $I=[2]\ne0$ for an invertible subalgebra~\cite{chen2026invertible} hosting it. 
\end{remark}

\begin{remark}[Invertible quasi-local algebra]\label{rem:infinite}
In~\cite{jones2026quantum}, invertible quasi-local algebras with infinite-dimensional local von Neumann algebras were introduced. The invariant considered in this article vanishes for these examples, since the GNVW index is trivial in this setting.
\end{remark}

\bibliographystyle{alpha}
\bibliography{refs}

@article{Haah-ISA,
  author = {Haah, Jeongwan},
  title = {Invertible subalgebras},
  journal = {Comm. Math. Phys.},
  volume = {403},
  pages = {661--698},
  year = {2023},
  note = {arXiv:2211.02086}
}

@article{JY-spectrum,
  author = {Ji, Mattie and Yang, Bowen},
  title = {Quantum cellular automata: the group, the space, and the spectrum},
  year = {2026},
  note = {arXiv:2602.16572}
}

@article{JY-obstruction,
  author = {Ji, Mattie and Yang, Bowen},
  title = {{$K$}-theoretic obstructions to linearizing {QCA} representations},
  year = {2026},
  note = {arXiv:2606.19657}
}

@article{HFH,
  author = {Haah, Jeongwan and Fidkowski, Lukasz and Hastings, Matthew B.},
  title = {Nontrivial quantum cellular automata in higher dimensions},
  journal = {Comm. Math. Phys.},
  volume = {398},
  pages = {469--540},
  year = {2023}
}

@article{Haah-Clifford,
  author = {Haah, Jeongwan},
  title = {Clifford quantum cellular automata: trivial group in 2d and {W}itt group in 3d},
  journal = {J. Math. Phys.},
  volume = {62},
  number = {9},
  year = {2021}
}

@article{Haah-Clifford-cat,
  author = {Haah, Jeongwan},
  title = {Topological phases of unitary dynamics: classification in {C}lifford category},
  journal = {Comm. Math. Phys.},
  volume = {406},
  number = {4},
  pages = {76},
  year = {2025}
}

@article{FH,
  author = {Freedman, Michael and Hastings, Matthew B.},
  title = {Classification of quantum cellular automata},
  journal = {Comm. Math. Phys.},
  volume = {376},
  pages = {1171--1222},
  year = {2020}
}

@article{GNVW,
  author = {Gross, D. and Nesme, V. and Vogts, H. and Werner, R. F.},
  title = {Index theory of one dimensional quantum walks and cellular automata},
  journal = {Comm. Math. Phys.},
  volume = {310},
  pages = {419--454},
  year = {2012}
}

@article{SYWTC,
  author = {Sun, Meng and Yang, Bowen and Wang, Zongyuan and Tantivasadakarn, Nathanan and Chen, Yu-An},
  title = {Clifford quantum cellular automata from topological quantum field theories and invertible subalgebras},
  journal = {PRX Quantum},
  volume = {7},
  pages = {010362},
  year = {2026},
  note = {arXiv:2509.07099}
}

@article{SZJL,
  author = {Shirley, Wilbur and Zhang, Carolyn and Ji, Wenjie and Levin, Michael},
  title = {Anomaly-free symmetries with obstructions to gauging and onsiteability},
  journal = {Phys. Rev. Lett.},
  volume = {136},
  pages = {216602},
  year = {2026},
  note = {arXiv:2507.21267}
}

@misc{Haah2026FermionicQCATalk,
  author       = {Haah, Jeongwan},
  title        = {Fermionic {QCA} in {2d} are trivial},
  year         = {2026},
  month        = mar,
  howpublished = {Talk at the CMSA Freedman Seminar, Harvard University},
  url          = {https://www.math.harvard.edu/event/quantum-cellular-automata-via-algebraic-k-theory-fermionic-qca-in-2d-are-trivial/}
}

@article{kwan2026fermionic,
  title={Fermionic quantum cellular automata in 2d are trivial},
  author={Kwan, Jeffrey and Long, David M and Haah, Jeongwan},
  journal={arXiv preprint arXiv:2609.09317},
  year={2026}
}

@article{yang2026categorifying,
  title={Categorifying {Clifford} {QCA}},
  author={Yang, Bowen},
  journal={Communications in Mathematical Physics},
  volume={407},
  number={4},
  pages={77},
  year={2026},
  publisher={Springer}
}

@article{sun2026non,
  title={Non-Clifford quantum cellular automata from invertible topological quantum field theories},
  author={Sun, Meng and Wang, Zongyuan and Yang, Bowen and Tantivasadakarn, Nathanan and Chen, Yu-An},
  journal={arXiv preprint arXiv:2607.21697},
  year={2026}
}

@article{shirley2022three,
  title={Three-dimensional quantum cellular automata from chiral semion surface topological order and beyond},
  author={Shirley, Wilbur and Chen, Yu-An and Dua, Arpit and Ellison, Tyler D and Tantivasadakarn, Nathanan and Williamson, Dominic J},
  journal={PRX quantum},
  volume={3},
  number={3},
  pages={030326},
  year={2022},
  publisher={APS}
}

@article{chen2026invertible,
  title={Invertible subalgebras for chiral semion topological order},
  author={Chen, Yu-An and Sun, Meng and Tantivasadakarn, Nathanan and Wang, Zongyuan and Yang, Bowen},
  journal={arXiv preprint arXiv:2610.18953},
  year={2026}
}

@article{fidkowski2024pumping,
  title={Pumping chirality in three dimensions},
  author={Fidkowski, Lukasz and Hastings, Matthew B},
  journal={Physical Review B},
  volume={109},
  number={23},
  pages={235142},
  year={2024},
  publisher={APS}
}

@article{fidkowski2025quantum,
  title={A quantum cellular automaton for every symmetry protected topological phase},
  author={Fidkowski, Lukasz and Haah, Jeongwan and Hastings, Matthew B},
  journal={Physical Review B},
  volume={112},
  number={3},
  pages={035123},
  year={2025},
  publisher={APS}
}

@article{jones2026quantum,
  title={Quantum cellular automata and invertible phases of matter},
  author={Jones, Corey and Sopenko, Nikita and Thorngren, Ryan},
  journal={arXiv preprint arXiv:2608.26456},
  year={2026}
}

@article{sopenko2024index,
  title={An index for invertible phases of two-dimensional quantum spin systems},
  author={Sopenko, Nikita},
  journal={arXiv preprint arXiv:2410.02059},
  year={2024}
}

@misc{Ludewig,
  author = {Ludewig, Matthias},
  title  = {Quantum cellular automata are a coarse homology theory},
  year   = {2026},
  note   = {arXiv:2603.10501}
}
\end{document}